\documentclass[aps,pra,floatfix,twocolumn,showpacs,10pt,longbibliography]{revtex4-2}
\usepackage{amsmath}
\usepackage{url}
\usepackage{graphicx}
\usepackage{dcolumn}
\usepackage{bm}
\usepackage{amssymb}
\usepackage{rotating}
\usepackage[abs]{overpic}
\usepackage[dvipsnames]{xcolor}
\usepackage{tabularx}
\usepackage{floatrow}
\usepackage{subfig} 
\usepackage[justification=RaggedRight]{caption}
\usepackage{hyperref}
\hypersetup{colorlinks = true, citebordercolor={black}, linkcolor={black}, citecolor={black}, urlcolor={black}}
\begin{document}








\title{\color{black} Cooper-Pair Condensates with Non-Classical Long-Range Order on Quantum Devices \color{black}}


\author{LeeAnn M. Sager and David A. Mazziotti}

\email{damazz@uchicago.edu}

\affiliation{Department of Chemistry and The James Franck Institute, The University of Chicago, Chicago, IL 60637}%

\date{Submitted June 21, 2021\textcolor{black}{; Revised August 23, 2021}\textcolor{black}{; Revised October 19, 2021}}



\begin{abstract}
An important problem in quantum information is the practical demonstration of non-classical long-range order on quantum computers.   One of the best known examples of a quantum system with non-classical long-range order is a superconductor.  Here we achieve \color{black} Cooper-like \color{black} pairing of qubits on a quantum computer\color{black}, which can be interpreted as superconducting or superfluid states via a Jordan-Wigner mapping.  \color{black}  We rigorously confirm the quantum long-range order by measuring the large $O(N)$ eigenvalue of the two-electron reduced density matrix.  The demonstration of maximal quantum long-range order is an important step towards more complex modeling of \color{black} phenomena with significant quantum long-range order on quantum computers such as superconductivity and superfluidity. \color{black}
\end{abstract}




\maketitle

\section{Introduction}

Phenomena like superconductivity and superfluidity arise from a Bose-Einstein-like condensation of fermion pairs into a quantum state with large non-classical long-range order~\cite{BCS1957, Blatt_SC, Anderson_2013, Drozdov_250, Ginzburg_1991, crabtree_2020, twist_2018, twist_2020_1, twist_2020_2,TSH2004, Fil_Shevchenko_Rev, Shiva, Kogar2017, LWT2017,london1950superfluids,FEYNMAN195517,leggett_1999,gorter_2011,guo_2020,hao_2020,delpace_2021}.   \color{black} Recently, quantum computers have emerged as potentially powerful calculators of \color{black} correlated quantum systems~\cite{verstraete_2009, Roushan_2017, smith_2019, mooney_2019, schuster_2019, huang_2020, mcardle_2020,HeadMarsden2021, Smart2019a, Sager_2020,Bharti_2021},
\color{black}
which foreshadows the potential emergence of a significant advantage of quantum computers over classical computers for certain classes of problems---a phenomenon known as quantum advantage \cite{arute_2019,elfving_2020,Wu_2021}.  \color{black}  Here, we  prepare and measure \color{black} Cooper-like pairs of qubit particles on a transmon-qubit quantum computer.  As the distinction between bosonic and fermionic statistics is lost as a result of a Jordan-Wigner mapping, condensations of Cooper-like qubit pairs can be interpreted as fermion-pair condensations, which can represent superconducting (or superfluid) states.

\begin{figure}[tbh!]
\begin{center}
\includegraphics[width=9cm]{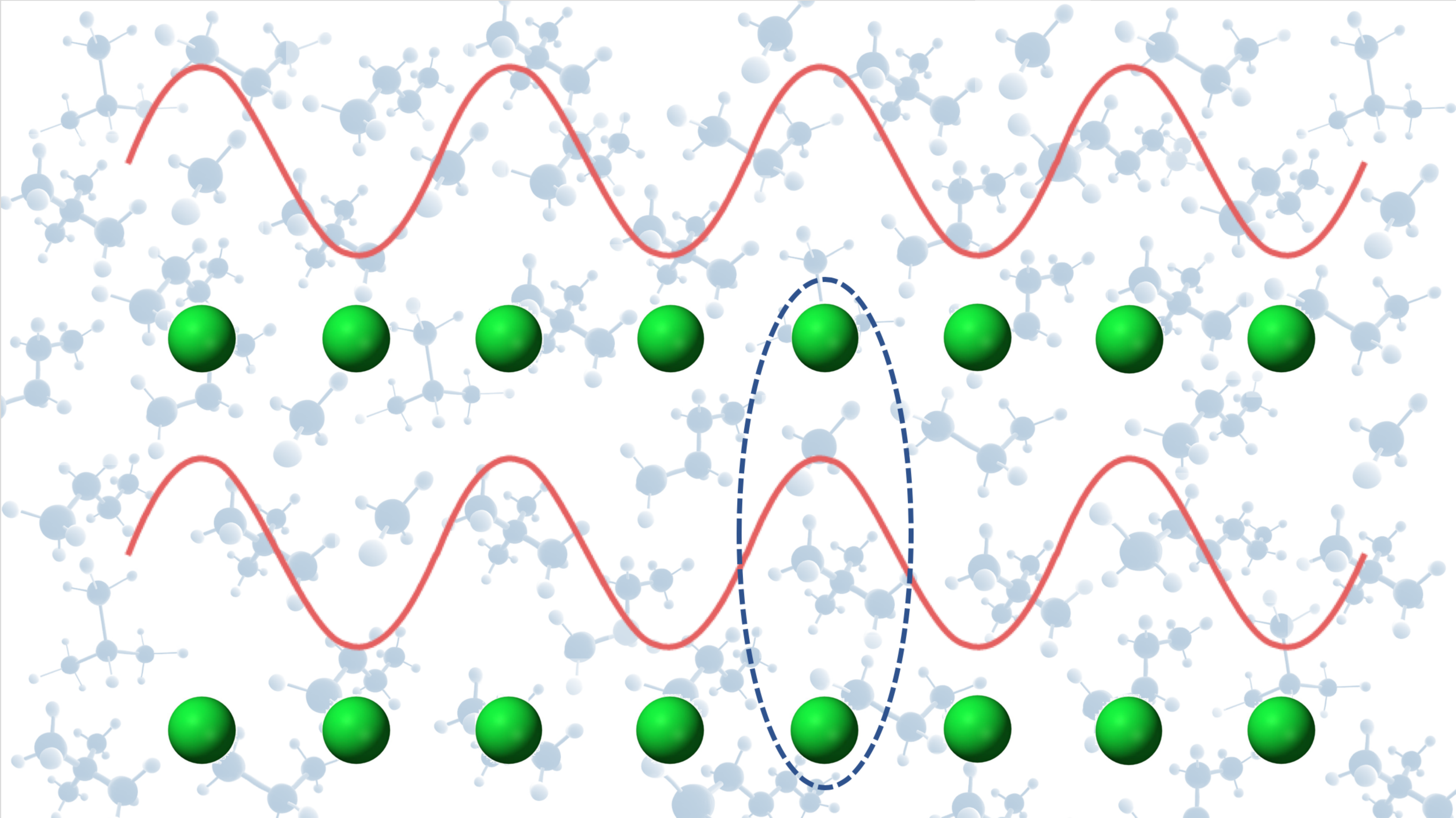}
\end{center}
\caption{\label{fig:PairedFig} A schematic demonstrating the interpretation of the Cooper pairing of \color{black} qubit particles \color{black} to create an overall \color{black} Cooper-like paired \color{black} state in a quantum system.}
\end{figure}

\begin{figure}[tbh!]
\begin{center}
\includegraphics[width=9cm]{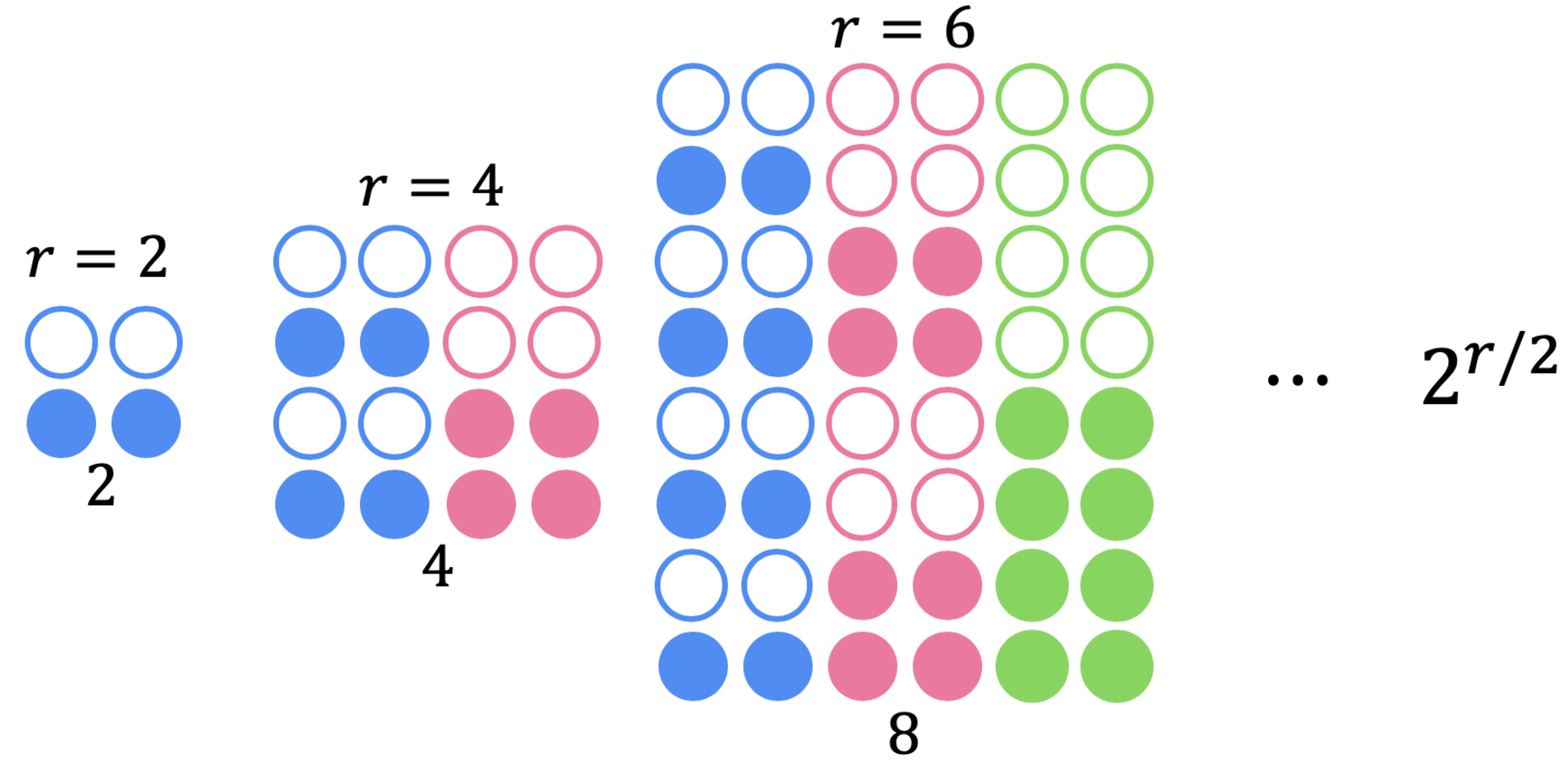}
\end{center}
\caption{A schematic demonstrating the possible configurations (i.e., each row) for a given number $r$ of qubits where a filled circle indicates the $|1\rangle$ state which corresponds to an occupied orbital and an unfilled circle represents the $|0\rangle$ state which corresponds to an unoccupied orbital. }
\label{fig:Configurations}
\end{figure}

In this study, qubit particles---which are hard-core bosons---\color{black} are entangled into Cooper-like bosonic pairs (see Fig. \ref{fig:PairedFig}) to form superconducting-like states---extreme antisymmetrized geminal power (AGP) wave functions~\cite{Penrose_BEC, Y1962, S1965, Shiva,C1963, C1964,Coleman_1965, sager_2019, Aly_2015,blatt_1964,bloch_1965,khamoshi_2021}.   As originally shown by Yang~\cite{Y1962} and Coleman~\cite{C1963,Coleman_1965,coleman_1989}, such states are extreme in the set of two-electron reduced density matrices (2-RDM), exhibiting the largest possible eigenvalue of the 2-RDM on the order of the number $N$ of electrons $O(N)$ that represents the maximum possible number of Cooper pairs in a common two-electron (geminal) eigenfunction of the 2-RDM.  We use tomography on the quantum computer to measure \color{black} a \color{black} subblock of the 2-RDM \cite{Coleman_1965,Kade_2017,Poelmans_2015} \color{black}(see Eq. (\ref{eq:d2})) \color{black} containing the large eigenvalue.  Diagonalization of this subblock on a classical computer produces the large eigenvalue and confirms the preparation of the extreme states with maximal non-classical (off-diagonal) long-range order \color{black} (ODLRO) \color{black} \cite{Y1962,coleman_1989}.  Even though the extreme AGP functions are expressible as projections of product wave functions \cite{C1963,Coleman_1965}, they have contributions from an exponentially scaling number of  orbital-product configurations (see Fig. \ref{fig:Configurations}).  Moreover, the measurement of the large eigenvalue of the 2-RDM is applicable to confirming non-classical long-range order in a much richer set of quantum states.  \color{black}
Because a necessary criterion for the modeling of superconductors (superfluids) on the quantum computer is the ability to capture the ODLRO, its demonstration provides a first step towards modeling more-complex superconducting (superfluid) materials.
\color{black}

\color{black}

\section{Theory}
\subsection{The Wavefunction:}
The \color{black} superconducting-like \color{black} state on the quantum computer is prepared by entangling pairs of qubits into Cooper-like bosonic states.  Consider the creation of a state with a Cooper pair of electrons, an extreme geminal \cite{Blatt_SC,C1963,C1964,Coleman_1965}, from the vacuum state
\begin{equation}
\label{eq:gf}
| g \rangle = \sum_{j}{e^{i\theta} {\hat a}^{\dagger}_{{\bar j}} {\hat a}^{\dagger}_{j}} | \varnothing \rangle
\end{equation}
where $j$ and ${\bar j}$ are the indices of the paired orbitals $\phi_{j}$ and $\phi_{{\bar j}}$, the sum over $j$ is taken with respect to all pairs, and $\theta$ is an arbitrary global phase.  If we represent each orbital by a qubit with the $|0\rangle$ state representing an unfilled orbital and the $|1\rangle$ state representing a filled orbital, we can use \color{black} a specific Klein transformation \cite{Klein_1938} known as the \color{black} Jordan-Wigner mapping \cite{jordan_wigner_1928}
\begin{equation}
{\hat a}^{\dagger}_{j} = e^{i\pi \sum_{k=1}^{j-1}{\sigma^{+}_{k} \sigma^{-}_{k}}}\sigma^{+}_{j}
\end{equation}
to map the fermionic operators in Eq.~(\ref{eq:gf}) to qubit operators to obtain
\begin{equation}
\label{eq:gq}
| g \rangle =  \sum_{j}{e^{i\theta} e^{i\pi \sum_{k=1}^{{\bar j}-1}{\sigma^{+}_{k} \sigma^{-}_{k}}}\sigma^{+}_{{\bar j}} e^{i\pi \sum_{k=1}^{j-1}{\sigma^{+}_{k} \sigma^{-}_{k}}}\sigma^{+}_{j}} | \varnothing \rangle .
\end{equation}
If the paired orbital indices $j$ and ${\bar j}$ are selected to be consecutive integers in the range $[1,r]$ where $r$ is the total number of orbitals, then the \color{black} Jordan-Wigner mappings \color{black} in  Eq.~(\ref{eq:gq}) simplify to a negative global phase which we can cancel by selecting $\theta=\pi$ to obtain
\begin{equation}
\label{eq:gq2}
| g \rangle =   \sum_{j}{\sigma^{+}_{{\bar j}} \sigma^{+}_{j}} | \varnothing \rangle .
\end{equation}
Hence, the extreme geminal of the Cooper pair $| g_{j{\bar j}} \rangle$ of electrons can be represented as two-qubit excitations without approximation.  The difference between the fermion and qubit statistics, typically included through an explicit many-qubit \color{black} Jordan-Wigner mapping \color{black}, disappears from the pairing of the orbitals to generate an extreme geminal.  Moreover, the explicit details of the pairing of the particles is contained within the unspecified orbitals $\phi_{j}$ and $\phi_{\bar j}$.  Consequently, the extreme geminal can physically represent Cooper pairing of electrons in a superconductor or a superfluid in addition to representing even the Cooper-like pairing of qubit particles \color{black}(hard-core \color{black} bosons) \cite{Keilmann_2008} which are paraparticles \cite{Wu_2002,mazz_mazz_2021}.

\begin{figure}[tbh!]
\begin{center}
\includegraphics[width=4cm]{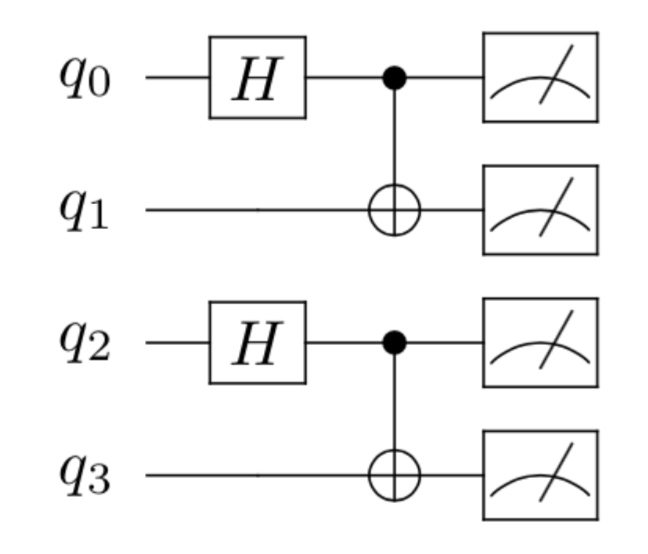}
\end{center}
\caption{\color{black} A schematic demonstrating the quantum state preparation for the $r=4$ AGP wavefunction given in Eq. (\ref{eq:QuantumState}) where $H$ represents the Hadamard gate which maps $|0\rangle$ to $(|0\rangle+|1\rangle)/\sqrt{2}$ and $|1\rangle$ to $(|0\rangle-|1\rangle)/\sqrt{2}$ and where each pair of qubits is connected via a standard controlled-NOT (CNOT) gate with control $\bullet$ and target $\oplus$.}
\label{fig:N4_circuit}
\end{figure}

The $N$-electron extreme AGP wave function $|\Psi_{\rm AGP}^{N} \rangle$ for even $N$ can be generated from the wedge product of the extreme geminal with itself $N/2$ times \cite{C1963,C1964,Coleman_1965,mazz_2011}
\begin{equation}
|\Psi_{\rm AGP}^{N} \rangle = | g(12) \rangle \wedge | g(34) \rangle  \wedge ...  \wedge | g((N-1)N) \rangle
\end{equation}
where the wedge $\wedge$ denotes the sum of all products resulting from the antisymmetric permutation of the particles.  We can also consider a wave function $|\Psi_{\rm AGP} \rangle$, also known as a Bardeen-Cooper-Schrieffer (BCS) wave function \cite{BCS1957}, that is a linear combination of the  $|\Psi_{\rm AGP}^{N} \rangle$ for all $N$ which is expressible as a product state
\begin{equation}
\label{eq:BCS}
| \Psi_{\rm AGP} \rangle = \prod_{j=1}^{r/2}{(1+e^{i\theta} {\hat a}^{\dagger}_{2j} {\hat a}^{\dagger}_{2j-1})} | \varnothing  \rangle .
\end{equation}
Using the Jordan-Wigner transformation and simplifying as above, we can generate the AGP state in Eq.~(\ref{eq:BCS}) with the qubit excitation operators
\begin{equation}
| \Psi_{\rm AGP} \rangle = \prod_{j=1}^{r/2}{(1+ {\hat \sigma}^{+}_{2j} {\hat \sigma}^{+}_{2j-1})} | \varnothing \rangle ,
\end{equation}
which can also be cast as the tensor multiplication of $r/2$ distinct extreme geminals (or the $|\Phi^+\rangle$ Bell states \cite{nielsen_2010})
\begin{equation}
|\Psi_{\rm AGP} \rangle=\mathop{\otimes}\limits_{j=1}^{r/2}\frac{1}{\sqrt{2}} \left[|00\rangle_{2j,2j-1}+|11\rangle_{2j,2j-1}\right]
\label{eq:QuantumState}
\end{equation}
where $j$ specifies the pair index and adjacent qubits with qubit indices $2j-1$ and $2j$ paired by definition. \color{black} This state, which is composed of substates with all possible, paired, even-numbered excitations, can be prepared on a quantum device according to the general gate sequence given in Eq.~(\ref{eq:overallprep}).  Figure~\ref{fig:N4_circuit} shows the specific $r=4$ preparation.

\color{black}
On the quantum computer, the extreme AGP state is physically composed of Cooper-like pairs of qubits.  Because the phase changes from the fermionic statistics are lost in the pairing---as seen in the above fermionic encoding of the qubits, the extreme state rigorously represents not only entangled pairs of qubits but also Cooper pairs of electrons that are entangled in superconducting or superfluid states.   Moreover, the state can represent any physical model for pairing as the precise nature of the pairing (i.e. the pairing of electrons in physical space or in momentum space) is contained in the paired orbitals $\phi_{j}$ and $\phi_{\bar j}$, which are left unspecified.  All pairing states have fundamental entanglement and order properties that are independent of the physical definition of the orbitals.  The non-classical long-range order of the extreme AGP state can be assessed from the number of Cooper pairs in the same extreme geminal, which is determinable from the largest eigenvalue of the 2-RDM \cite{Y1962,S1965,C1963}.

\subsection{The Signature of \color{black} Non-Classical Off-Diagonal Long-Range Order (ODLRO) \color{black}}
In order to measure whether the experimentally-prepared quantum state and/or the number-conserving substates---\color{black}which are all possible, even eigenstates of the number operator (Eq. (\ref{eq:N})) that can be projected out from the overall ensemble quantum state according to the methodology presented in  \ref{sec:numbercons}---demonstrate off-diagonal long-range order\color{black}, we conduct quantum tomography \color{black} (see \ref{sec:tomographyfull} and \ref{sec:numbercons}) \color{black} to probe directly the presence and extent of \color{black} ODLRO \color{black}. To determine the presence and degree of this long-range order for a specified quantum state, it is useful to establish a calculable, characteristic property \cite{Penrose_BEC, Y1962, S1965, Shiva, C1963, Coleman_1965, sager_2019, Aly_2015}.  Such a signature of \color{black} ODLRO \color{black} is a large eigenvalue in the 2-RDM, which we denote as $\lambda_D$ \cite{Y1962, S1965}.  \color{black} As this large eigenvalue corresponds to the number of Cooper-like qubit pairs occupying the same two-qubit geminal (which is directly analogous to the one-qubit orbital), any $\lambda_D$ value exceeding the Pauli-like limit of one is indicative of ODLRO \cite{Y1962,S1965,C1963}. \color{black}

While the entire 2-RDM can be measured by quantum tomography, only the following subblock of the 2-RDM \cite{Coleman_1965,Kade_2017,Poelmans_2015} is required due to the block diagonal structure of the 2-RDM of the AGP wave function
\begin{equation}
\label{eq:d2}
^{2} D^{j{\bar j}}_{k {\bar k}} = \langle \Psi_{\rm AGP} | {\hat a}^{\dagger}_{j} {\hat a}^{\dagger}_{\bar j}{\hat a}^{}_{\bar k} {\hat a}^{}_{k} | \Psi_{\rm AGP} \rangle
\end{equation}
\color{black}\noindent where $j/{\bar j}$ and $k/{\bar k}$ represent paired fermions, which are given by $j=2m/{\bar j}=2m-1$ and $k=2n/{\bar k}=2n-1$ for integers $m,n$ in the framework of the AGP wavefunction\color{black}.  After Jordan-Wigner transformation to the qubit representation, we can equivalently represent this block of the 2-RDM in terms of the qubit excitation operators as
\begin{equation}
^{2} D^{j{\bar j}}_{k {\bar k}} = \langle \Psi_{\rm AGP} | {\hat \sigma}^{\dagger}_{j} {\hat \sigma}^{\dagger}_{\bar j}{\hat \sigma}^{}_{\bar k} {\hat \sigma}^{}_{k} | \Psi_{\rm AGP} \rangle .
\end{equation}
For fixed number $N$ of electrons, if the 2-RDM is normalized to $N(N-1)$ as in second quantization, the maximum eigenvalue for even $N$ is bounded from above by $N$ as shown by Yang \cite{Y1962} and Sasaki \cite{S1965}.  Moreover, for a finite rank of $r$ orbitals, this bound can be further tightened \cite{C1963,Coleman_1965} to
\begin{equation}
\label{eq:bnd}
\lambda_{D} \le N \left ( 1-\frac{N-2}{r} \right ) .
\end{equation}
While the thermodynamic limit is not reached until $r \rightarrow \infty$, even for finite $r$, as long as $N \ge 4$, the 2-RDM exhibits a large eigenvalue that represents the non-classical long-range order associated with Cooper pairing.  The 2-RDM from the non-number conserving extreme AGP state $| \Psi_{\rm AGP} \rangle$ also exhibits a large eigenvalue, representing an average of the Cooper pairs in each of the fixed-$N$ extreme AGP states \color{black}(i.e., the number-conserving substates)\color{black}.  The number-conserving blocks of the 2-RDM with even particle numbers---i.e., 2-RDMs of zero,  two, four, ..., $r-2$, and $r$ particles for an $r$-qubit system---can be determined from the non-number conserving state via post-measurement analysis \color{black}(see \ref{sec:numbercons})\color{black}.  Analysis on the presence and extent of \color{black} ODLRO\color{black} (measured via $\lambda_D$) of both the overall entangled state ($|\Psi\rangle$) and the number-conserving substates is conducted for various numbers $r$ of total qubits in the following sections.

\section{Results}
\begin{figure}[tbh!]
\begin{center}
\includegraphics[width=9cm]{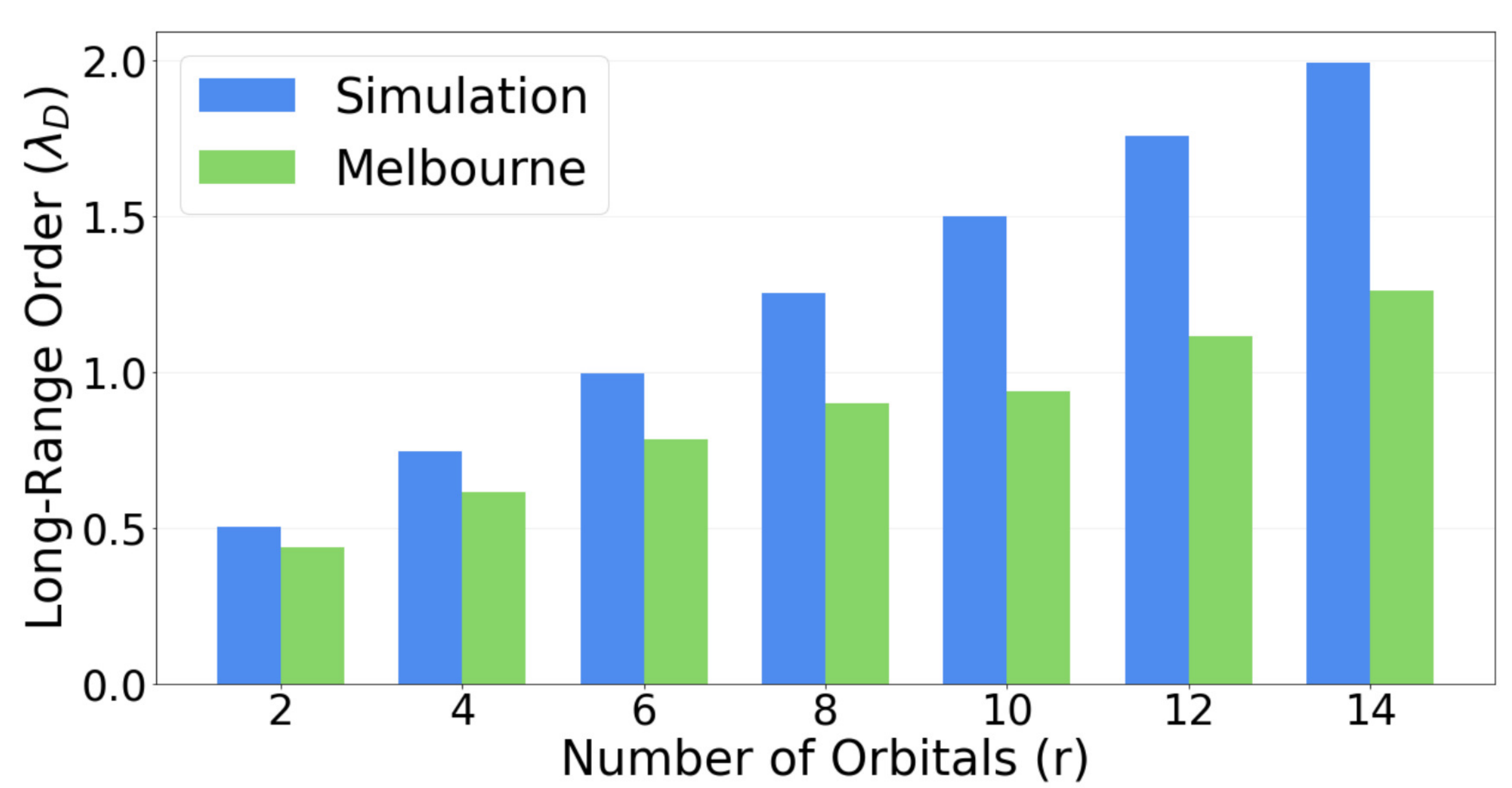}
\end{center}
\caption{ \label{fig:BarPlot1} The $\lambda_D$ values for the overall ensemble state preparation for \color{black} QASM \color{black} simulation and experimental melbourne results. }
\end{figure}

The extreme non-number conserving AGP state is prepared for both \color{black} simulation utilizing IBM's QASM simulator (ibmq\_qasm\_simulator) \color{black} and an experimental quantum device for all even-numbered qubit systems from $r=0$ to $r=14$. Post-measurement computation of the quantum signature of  off-diagonal long-range order  ($\lambda_D$) is then employed to probe the presence and extent of \color{black} ODLRO \color{black} for these overall states. As can be seen in Figure~\ref{fig:BarPlot1}, the signature of \color{black} ODLRO \color{black} increases as the number $r$ of qubits comprising the system is increased, and---for \color{black} QASM \color{black} simulation---\color{black} ODLRO \color{black} is observed (i.e., $\lambda_D > 1$) for all prepared states with $r \ge 8$. While the experimental results deviate from \color{black} QASM \color{black} simulation due to the noisy nature of near-term quantum devices \cite{preskill_2018} \color{black} (see \ref{sec:noise} and \ref{sec:devicedetails})\color{black}, experimental systems with $r=12$ and $r=14$ qubits did demonstrate \color{black} ODLRO\color{black}. Further, the trend of the extent of \color{black} ODLRO \color{black} increasing as the number of qubits comprising the system increases holds for the experimental results, which is promising for future benchmarking of quantum computers through the preparation of extreme \color{black} AGP \color{black} states with larger number of qubits as well as efforts to probe more macroscopically-scaled \color{black} materials demonstrating ODLRO \color{black} on quantum devices.

\color{black}
As the non-number conserving extreme AGP state is an ensemble composed of number-conserving substates, the large eigenvalue associated with the ODLRO of the ensemble state is the ensemble average of the substates.  By definition, then, the long-range order of the ensemble is less than that of the substate with the largest degree of ODLRO, which is expected to occur around the center of the number distribution $N \approx r/2$.
\color{black}
Additionally, real-world \color{black} materials demonstrating ODLRO such as superconductors \color{black}  should conserve particle number. It is hence beneficial to probe the number-conserving substates that comprise the overall entangled state in order to both isolate the \color{black} ODLRO \color{black} behavior of the number-conserving substates and to more-closely model real-world \color{black} materials\color{black}.

\begin{figure}[tbh!]
\begin{center}
\includegraphics[width=9cm]{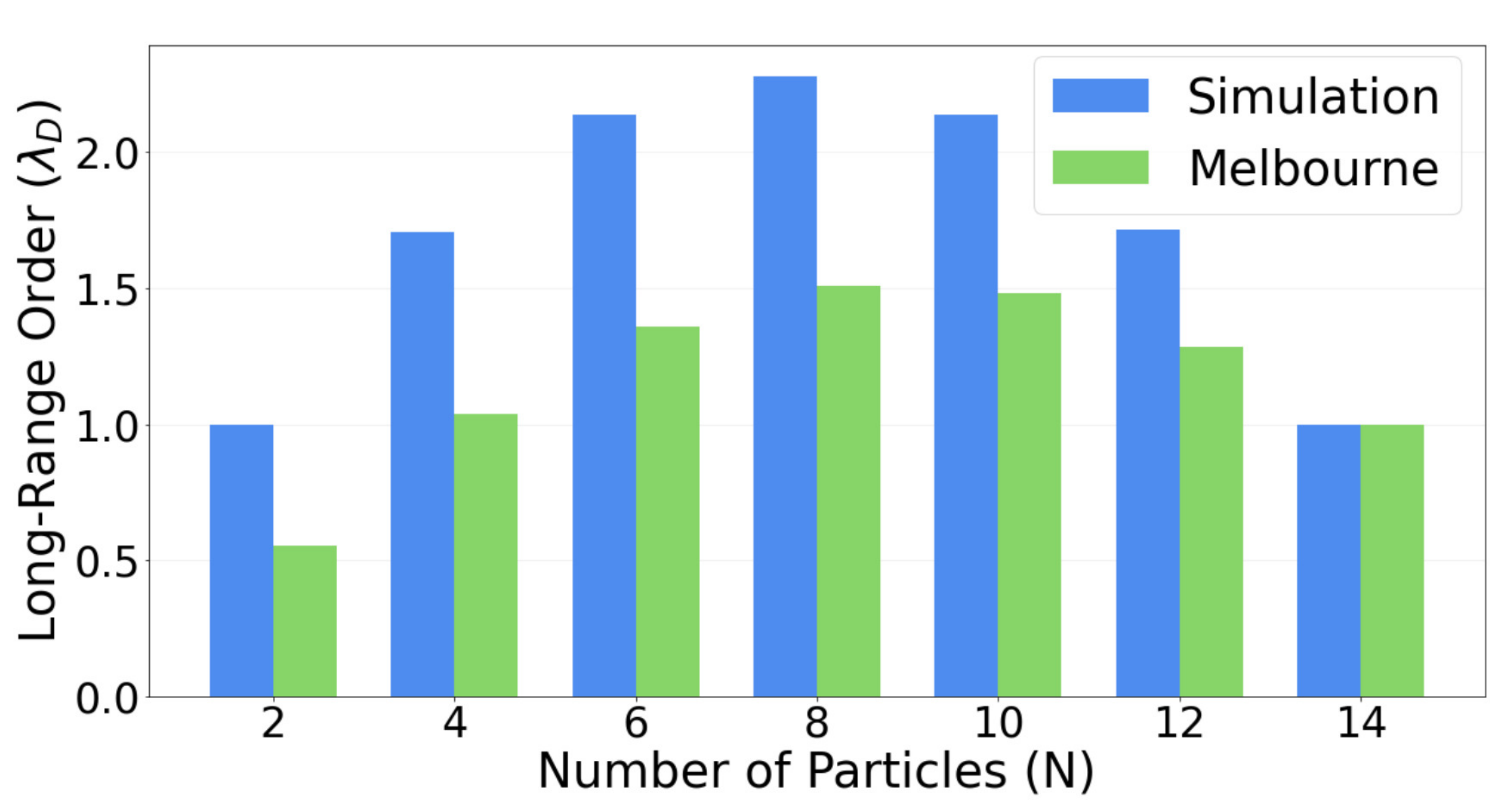}
\end{center}
\caption{ \label{fig:BarPlot2} The $\lambda_D$ values for the number-conserving substates of rank $r=14$ for \color{black} QASM \color{black} simulation and experimental melbourne results. }
\end{figure}

By projecting out a specific number of particles from the results obtained for overall entangled state \color{black} (see \ref{sec:numbercons})\color{black}, we can probe the behavior and properties of the number-conserving substates. Specifically, as is shown in Fig.~\ref{fig:BarPlot2}, the extent of \color{black} ODLRO \color{black} ($\lambda_D$) for each number-conserving state can be isolated from the overall $r$-qubit preparation described in Eq.~(\ref{eq:QuantumState}). As can be seen from the \color{black} QASM \color{black} simulation results, all number-conserving substates with $2<N<r$ demonstrate \color{black} ODLRO \color{black} ($\lambda_D>1$) where $N=2$ fails to demonstrate condensation behavior as the maximum signature of condensation is $N/2$ for even $N$-particle systems~\cite{Y1962,S1965} and where $N=r$ fails to demonstrate condensation behavior as this
\color{black} substate describes the state in which all orbitals are fully occupied with no entanglement---i.e., the single Slater determinant $|1\rangle^{\otimes r}$---and is hence expected to have a maximum eigenvalue of $\lambda_D=1$.
\color{black}
Further, the signature of condensation seems to follow a bell curve centered around $(r+2)/2$ such that maximum \color{black} ODLRO \color{black} is observed at half filling for $N=(r+2)/2$ if $(r+2)/2$ is even and for both $N=(r+2)/2-1$ and $(r+2)/2+1$ if $(r+2)/2$ is odd.  Again, the extent of \color{black} ODLRO \color{black} is lesser for the experimental results for all particle-conserving states due to experimental error; however, the qualitative trends described for \color{black} QASM \color{black} simulation hold in general although the bell curve does demonstrate a slight negative (right-modal) skew, implying that the quantum computer does not exactly treat the particle and hole statistics symmetrically.  Importantly, \color{black} ODLRO \color{black} is clearly observed for $r=14$ experimental results for particle numbers $N\ge6$. Note that although only results for the largest-qubit preparation $r=14$ are shown, all data is included in \color{black} Table \ref{tab:mel}\color{black}; the trends in the $r=14$ data hold for the lower-qubit results, and additionally, the $r=14$ qubit data demonstrates the largest signature of \color{black} off-diagonal long-range order \color{black} as the largest  $\lambda_D$ value for a fixed $N$ increases as the number $r$ of qubits is increased.

\section{Conclusions}
Here we prepare superconducting\color{black}-like \color{black} states from the Cooper pairing of qubits on a  transmon quantum computer---where each qubit is composed of a microwave phonon in an anharmonic well potential.  Using the Jordan-Wigner mapping between fermions and qubits, we rigorously show that the prepared states are equally valid representations of condensations of Cooper pairs of fermions, bosons, or qubits.  Hence, such Cooper-pair-based condensations and their associated non-classical long-range order are independent of the particle statistics.  Moreover, the prepared states are also independent of the physical details of the paired orbitals, and consequently, are representative of superconducting, superfluid, or other pairing states.  The studied states are known as extreme AGPs because they exhibit the maximum degree of non-classical (off-diagonal) long-range order as determined by the number of Cooper pairs in the same geminal state, which is equal to the largest eigenvalue of the 2-RDM \cite{Y1962,S1965,blatt_1964,Blatt_SC,C1963,Coleman_1965,coleman_1989}.  We measure a subblock of the 2-RDM \cite{Coleman_1965,Kade_2017,Poelmans_2015} on the quantum computer and compute its largest eigenvalue on a classical computer.   We observe large eigenvalues both for the non-particle conserving extreme AGP state and the particle-conserving extreme AGP substates.  The large eigenvalues confirm the preparation of these extreme AGP states, which are the only states to exhibit the largest possible eigenvalues \cite{Coleman_1965}, as well as the generation of maximum non-classical long-range order.

The upper bound on the largest eigenvalue of the 2-RDM, \color{red} $\lambda_{D} = N$, \color{black} is technically only reached in the thermodynamic limit of $r \rightarrow \infty$.   However, as seen in Eq.~(\ref{eq:bnd}), the large eigenvalue is rapidly approached with increasing $r$ as the fraction of Cooper pairs that are removed from the condensate due to finite size effects scales as $1/r$.  Consequently, the quantum long-range order as well as its associated entanglement begin to appear for the range of $r$ ($r \le 14$) explored in the present study.  On both \color{black} IBM's QASM simulator \color{black} and an IBM quantum computer, we observe that the large eigenvalue follows the expected bell curve with respect to $r$.  The known value for the maximum eigenvalue of the extreme AGP state provides a clear metric for not only confirming the presence of the extreme state and its long-range order but also benchmarking the fidelity  with respect to noise of both current and future quantum computers.

\color{black} An aspirational goal of quantum computing is to achieve a quantum advantage over traditional classical computing for the solution of a significant problem.  One such area of chemistry and physics, which traces back to the original proposal of Feynman \cite{feynman_1982}, is the simulation of molecules on quantum computers.  The construction of the wave function on a classical computer scales exponentially in the number of orbital-based configurations.  In principle, the quantum computer offers the possibility of preparing and measuring quantum states with non-exponential scaling. \color{black}  In the present case, the extreme AGP wave function is a product state composed of a product of extreme geminals (or Bell states).  Consequently, the maximum degree of non-classical long-range order, at least as measured by the largest eigenvalue of the 2-RDM, can be achieved with polynomial cost on both classical and quantum computers.  The extreme AGP states, nonetheless, provide an intriguing reference for the exploration of more complicated \color{black} states demonstrating large ODLRO (i.e., demonstrating large eigenvalues) \color{black} that cannot be easily expressed as product-state wave functions.   From this perspective, the present work of preparing and measuring superconducting\color{black}-like \color{black} states from Cooper pairs of qubits on a quantum computer provides an initial step towards \color{black} investigating \color{black} more complicated condensates of Cooper pairs---with potential future applications to the study of both superconducting materials and simulation.

\color{black} Certain electronic structure methodologies---such as Hartree Fock (HF), density functional theory (DFT) and coupled cluster (CCSD) theory---are incapable of demonstrating off-diagonal long range order (ODLRO) indicative of particle-particle condensation.  As such, in order to accurately explore complex, real-world superconducting states on quantum devices, it is necessary to establish that noisy intermediate quantum devices are both capable of exhibiting ODLRO as well as the features of state preparations that yield such ODLRO.  In the strongly-correlated regime, HF-based and multi-reference approaches on quantum devices are not suitable starting points for state preparations for exploring ODLRO.  Instead, we show that number-projected BCS wavefunctions---alternatively termed AGP wavefunctions---are ideal starting points for building more-complicated wavefunctions that may model real-world superconducting states as we demonstrate that these AGP wavefunctions created by the pairing of adjacent qubits are capable of demonstrating ODLRO on the noisy near-term quantum devices.  Such insight may additionally prove useful in the development of an appropriate ansatz for hybrid quantum-classical algorithms---such as the variational quantum eigensolver (VQE)---in order to accurately describe strongly-correlated systems demonstrating ODLRO.  \color{black}

\begin{acknowledgments}
D.A.M. gratefully acknowledges the Department of Energy, Office of Basic Energy Sciences, Grant DE-SC0019215 and the U.S. National Science Foundation Grants No. CHE-2035876, No. DMR-2037783, and  No. CHE-1565638.
\end{acknowledgments}

\appendix
\section*{Appendix}
\addcontentsline{toc}{section}{Appendix}
\renewcommand{\thesubsection}{Appendix \Alph{subsection}}
We include details on the quantum algorithms used to prepare the qubit states presented in the article; the quantum tomography of the particle-particle reduced density matrix {\color{black} for both the overall non-number conserving state as well as number-conserving substates}; a description of noise on near-term quantum devices; an analysis of errors by comparing simulated and experimental joint probabilities of occupation; relevant details on the experimental quantum backends employed; \color{black} full results obtained from ibmq\_qasm\_simulation, ibmq\_16\_melbourne, ibmq\_5\_yorktown, ibmq\_santiago, and ibmq\_rochester; and information regarding the state preparation fidelity. \color{black}

\subsection{State Preparations}
The overall quantum state is composed of all pairwise even excitations of the $r/2$ individually-paired qubits. This preparation is accomplished by
\begin{equation}
|\Psi\rangle = \left[\prod\limits_{p=0}^{r/2-1}C_{2p}^{2p+1}H_{2p}\right]|0\rangle^{\otimes r}
\label{eq:overallprep}
\end{equation}
where $|0\rangle^{\otimes r}$ is the initial quantum state in which all qubits are in their ground state \color{black} (i.e., all orbitals are unoccupied), \color{black} $p$ represents the index of each of the possible $r/2$ adjacent, paired qubits, $H_i$ is the Hadamard gate acting on $Qi$, and $C_i^j$ is a CNOT gate with $Qi$ and $Qj$ acting as the control and target qubits, respectively. Application of the gate sequence given in Eq. (\ref{eq:overallprep})\color{black}---represented pictorially for $r=4$ in Fig. \ref{fig:N4_circuit}---\color{black} produces the AGP wavefunction described by Eq. (\ref{eq:QuantumState}).

\subsection{Quantum Tomography for the Particle-Particle RDM}
\label{sec:tomographyfull}
\color{black} While the full particle-particle RDM has \color{black} elements given by
\begin{equation}
^{2} D_{k,l}^{i,j} = \langle \Psi | {\hat a}^{\dagger}_i {\hat a}^{\dagger}_j {\hat a}_k {\hat a}_l  | \Psi \rangle
\label{eq:D2}
\end{equation}
\color{black} for all possible combinations of one-boson spin orbitals indexed by $i$, $j$, $k$, and $l$ in a finite basis state with rank $r$, the large eigenvalue, $\lambda_D$, is contained within the subblock of the 2-RDM given by \cite{C1963,Coleman_1965}
\color{black}
\begin{widetext}
\color{black}
\begin{equation}
	\begin{array}{c|cccc}
                                             &  { \hat{a}_{1}\hat{a}_{0}} &  { \hat{a}_{3}\hat{a}_{2}} & {\cdots}  &  { \hat{a}_{r-2}\hat{a}_{r-1}}\\ \hline
   { \hat{a}_{0}^\dagger\hat{a}^\dagger_{1}} &{ \hat{a}_{0}^\dagger\hat{a}^\dagger_{1}\hat{a}_{1}\hat{a}_{0}}  &{ \hat{a}_{0}^\dagger\hat{a}^\dagger_{1}\hat{a}_{3}\hat{a}_{2}}  &{\cdots}  &{ \hat{a}_{0}^\dagger\hat{a}^\dagger_{1}\hat{a}_{r-2}\hat{a}_{r-1}}  \\
   { \hat{a}_{2}^\dagger\hat{a}^\dagger_{3}} &{ \hat{a}_{2}^\dagger\hat{a}^\dagger_{3}\hat{a}_{1}\hat{a}_{0}}  &{ \hat{a}_{2}^\dagger\hat{a}^\dagger_{3}\hat{a}_{3}\hat{a}_{2}}  &{\cdots}  &{ \hat{a}_{2}^\dagger\hat{a}^\dagger_{3}\hat{a}_{r-2}\hat{a}_{r-1}}  \\
   {\vdots}                                  &{\vdots}  &{\vdots}  &{\ddots}  &{\vdots}  \\
   { \hat{a}_{r-2}^\dagger\hat{a}^\dagger_{r-1}} &{ \hat{a}_{r-2}^\dagger\hat{a}^\dagger_{r-1}\hat{a}_{1}\hat{a}_{0}}  &{ \hat{a}_{r-2}^\dagger\hat{a}^\dagger_{r-1}\hat{a}_{3}\hat{a}_{2}}  &{\cdots}  &{ \hat{a}_{r-2}^\dagger\hat{a}^\dagger_{r-1}\hat{a}_{r-2}\hat{a}_{r-1}}  \\
\end{array}
\label{eq:submat}
\end{equation}
\end{widetext}
\color{black}

\color{black}
\noindent and, hence, only this portion of the matrix is constructed.  \color{black}  Note that  $\hat{a}_i^\dagger$ and $\hat{a}_i$ are creation and annihilation operators for orbital $i$ (and thereby qubit $Qi$), which can be represented in matrix form as
\begin{equation}
\hat{a}^\dagger_i=\left(\begin{array}{cc}
  	 0  & 0 \\
   	 1 & 0
	\end{array}\right)_i
\label{eq:creation}
\end{equation}
and
\begin{equation}
\hat{a}_i=\left(\begin{array}{cc}
  	 0  & 1 \\
   	 0 & 0
	\end{array}\right)_i
\label{eq:annihilation}
\end{equation}
\color{black}
such that each creation operator creates a particle in an empty orbital $i$---takes a qubit $i$ from $|0\rangle$ to $|1\rangle$---and each annihilation operator kills a particle in a filled orbital $i$---takes a qubit $i$ from $|1\rangle$ to $|0\rangle$---where
\begin{equation}
    |0\rangle=\left(\begin{array}{c} 1 \\ 0 \end{array}\right)
\end{equation}
and
\begin{equation}
    |1\rangle=\left(\begin{array}{c} 0 \\ 1 \end{array}\right)
\end{equation}

After construction of the subblock of the 2-RDM corresponding to the large eigenvalue (${}^{2}D_{\rm s.b.}$), the signature of off-diagonal long-range order---$\lambda_D$---is then obtained by solving the eigenvalue equation
\begin{equation}
{}^{2}D_{\rm s.b.}v_i^D=\epsilon_i^D v_i^D
\end{equation}
with the signature corresponding to the largest $\epsilon_i^D$ value.

\color{black}
\color{black}
\subsubsection{Tomography via Pauli Expectation Values}
\label{sec:tomviaP1}
The subblock of the 2-RDM containing the large eigenvalue (i.e., Eq. (\ref{eq:submat})) can be obtained via translation of each of its elements into the bases of Pauli matrices, the expectation values of which can be directly probed on a quantum device.  Specifically, the creation and annihilation operators can be rewritten as
\begin{equation}
\hat{a}^\dagger_i=\frac{1}{2}\left(X_i-iY_i\right)
\end{equation}
and
\begin{equation}
\hat{a}_i=\frac{1}{2}\left(X_i+iY_i\right)
\end{equation}
with diagonal elements being given by
\begin{gather}
\hat{a}^\dagger_j\hat{a}^\dagger_{j+1}\hat{a}_{j+1}\hat{a}_{j}=(\hat{a}^\dagger_j\hat{a}_j)(\hat{a}^\dagger_{j+1}\hat{a}_{j+1})\nonumber\\=\frac{1}{16}\left(X_j-iY_j\right)\left(X_j+iY_j\right)\left(X_{j+1}-iY_{j+1}\right)\left(X_{j+1}+iY_{j+1}\right)\nonumber\\
=\frac{1}{4}\left(I_j-Z_j\right)\left(I_{j+1}-Z_{j+1}\right)\nonumber\\
=\frac{1}{4}\left(1-Z_j-Z_{j+1}+Z_jZ_{j+1}\right)
\end{gather}
and off-diagonal elements being given by
\begin{widetext}
\begin{gather}
\hat{a}^\dagger_j\hat{a}^\dagger_{j+1}\hat{a}_{k+1}\hat{a}_k=\frac{1}{16}\left(X_j-iY_j\right)\left(X_{j+1}-iY_{j+1}\right)\left(X_{k+1}+iY_{k+1}\right)\left(X_k+iY_k\right)
\nonumber\\
\frac{1}{16}(X_jX_{j+1}X_kX_{k+1}+iX_jX_{j+1}X_kY_{k+1}+iX_jX_{j+1}Y_kX_{k+1}-X_jX_{j+1}Y_kY_{k+1}
\nonumber\\
-iX_jY_{j+1}X_kX_{k+1}+X_jY_{j+1}X_kY_{k+1}+X_jY_{j+1}Y_kX_{k+1}+iX_jY_{j+1}Y_kY_{k+1}
\nonumber\\
-iY_jX_{j+1}X_kX_{k+1}+Y_jX_{j+1}X_kY_{k+1}+Y_jX_{j+1}Y_kX_{k+1}+iY_jX_{j+1}Y_kY_{k+1}
\nonumber\\
-Y_jY_{j+1}X_kX_{k+1}-iY_jY_{j+1}X_kY_{k+1}-iY_jY_{j+1}Y_kX_{k+1}+Y_jY_{j+1}Y_kY_{k+1})
\end{gather}
\end{widetext}
\noindent Therefore, each 2-RDM matrix element can be obtained by directly probing the expectation values of four-qubit tensor products of Pauli matrices.

As all wavefunctions prepared in this study are real, the 2-RDM should consist of only real-valued elements; hence, all imaginary components of 2-RDM elements should approach zero within a small range dictated by randomness inherent to quantum systems as well as error on the device employed.  Therefore, only eight of the sixteen four-qubit expectation values corresponding to real contributions to a given 2-RDM element are nonzero and hence necessary for the determination of the subblock of the 2-RDM; thus, to lower computational expense, only these real values are used to construct the 2-RDM subblock where tomography via Pauli expectation values is conducted (as is consistent with previous analysis conducted in a manner similar to that described in Ref. \onlinecite{Sager_2020}).

\subsubsection{Tomography via Direct Computation of the Wavefunction}
\label{sec:tompsi}
As can be observed from Eq. (\ref{eq:QuantumState}), the phase angle of all qubits in the AGP wavefunction are known to be uniformly zero.  As such, knowledge of the probabilities with which each of the possible $2^r$ basis states ($|\eta_i\rangle$) for an $r$-qubit calculation are sampled out of the 81,920 times (8192 per trial) a given state is prepared and probed  is sufficient information to completely construct the wavefunction ($|\Psi\rangle$).  Specifically, the wavefunction takes the form of a vector (using traditional qubit vector notation \cite{Qiskit_Multiple}) with each element of the wavefunction $|\Psi_i\rangle$ corresponding to the basis state $|\eta_i\rangle$ being given by
\begin{equation}
|\Psi_i\rangle=+\sqrt{p(\eta_i)}=+\sqrt{|\langle \eta_i|\Psi\rangle|^2}=+|\langle \eta_i|\Psi\rangle|
\end{equation}
which is the positive square root of the probability ($p(\eta_i)$) with which the corresponding qubit basis state is measured.

Each individual element of the matrix shown in Eq. (\ref{eq:submat}) is then computed from the appropriate expectation value $\langle\Psi|\hat{a}^\dagger_j\hat{a}^\dagger_{j+1}\hat{a}_{k+1}\hat{a}_k|\Psi\rangle$ for the wavefunction in vector form obtained for a given state preparation.  The operator $\hat{a}^\dagger_j\hat{a}^\dagger_{j+1}\hat{a}_{k+1}\hat{a}_k$ can be represented as a $2^r\times 2^r$ matrix according to
\begin{widetext}
\begin{equation}
\hat{a}^\dagger_j\hat{a}^\dagger_{j+1}\hat{a}_{k+1}\hat{a}_k=\left[\bigotimes\limits_{a=1}^{j-1}\left(\begin{array}{cc}1&0\\0&1\end{array}\right)_{a}\right] \otimes \left(\begin{array}{cc}0&0\\1&0\end{array}\right)_{j} \otimes \left(\begin{array}{cc}0&0\\1&0\end{array}\right)_{j+1}
\left[\bigotimes\limits_{b=j+2}^{k-1}\left(\begin{array}{cc}1&0\\0&1\end{array}\right)_{b}\right]\otimes \left(\begin{array}{cc}0&1\\0&0\end{array}\right)_{k} \otimes \left(\begin{array}{cc}0&1\\0&0\end{array}\right)_{k+1} \left[\bigotimes\limits_{c=k+2}^{r}\left(\begin{array}{cc}1&0\\0&1\end{array}\right)_{c}\right]
\end{equation}
\end{widetext}
which is the tensor product of the creation and annihilation operators in matrix form (Eqs. (\ref{eq:creation}) and (\ref{eq:annihilation})) corresponding to the appropriate qubits ($j,j+1,k,k+1$) and identity matrices for each spectator qubit.  The expectation value can thus be computed directly via Matrix-Vector multiplication.

Comparing this method for the tomography of the subblock of the 2-RDM corresponding to the large eigenvalue (${}^{2}D_{\rm s.b.}^{\Psi}[i,j]$) to the tomography obtained via the expectation values of the Pauli matrices (${}^{2}D_{\rm s.b.}^{P}[i,j]$) yields Euclidean distances between the two matrices---represented mathematically as
\begin{equation}
\sqrt{\sum\limits_{i=0}^{r/2-1}\sum\limits_{j=0}^{r/2-1}\left({}^{2}D_{\rm s.b.}^{P}[i,j]-{}^{2}D_{\rm s.b.}^{\Psi}[i,j]\right)^2}
\end{equation}
---of $[r=2:0.001, r=4:0.004, r=6:0.003, r=8:0.005, r=10:0.007, r=12:0.009]$ for IBMQ QASM simulation, which indicates that any difference between the two methods is caused by inherent randomness in the absence of error.  Indeed, if \color{black} the QASM simulator is used, no difference is observed between the two methodologies.  On the error-prone melbourne experimental device, the Euclidean distances for the $r=2$ and the $r=8$ calculations are $[r=2:0.015, r=8:0.338]$, which is likely due to a large degree of error on the quantum device, which increases with the number of qubits (and hence the number of two-qubit gates) involved.  In fact, for the same number of qubits, the Euclidean distances between ${}^{2}D_{\rm s.b.}^{P}[i,j]$ and the ideal, expected ${}^{2}D_{\rm s.b.}$ matrix is $[r=2:0.063, r=8:0.541]$ and that between ${}^{2}D_{\rm s.b.}^{\Psi}[i,j]$ and ${}^{2}D_{\rm s.b.}$ is $[r=2:0.048, r=8:0.372]$, showing that the Euclidean distance between the two is on the same order or significantly smaller than the distance between each and the ideal.  If anything, the direct wavefunction methodology seems to provide a slight mitigation of errors, likely due to the smaller number of circuits run as only one circuit per trial for the direct wavefunction is necessary while multiple circuits per trial must be run in order to obtain the expectation values of the Pauli matrices.

Note that to decrease computational expense---especially in terms of the isolation of the number-conserving substates as explained in the following section---in this paper, we implement the tomography via the direct computation of the wavefunction.

\color{black}
\color{black}
\subsection{Isolation of Number-Conserving Components}
\label{sec:numbercons}
The number-conserving substates are eigenfunctions of the number operator,
\begin{equation}
\hat{N}=\sum\limits_{q=0}^{N-1}\hat{a}^\dagger_{q}\hat{a}_q=\sum\limits_{q=0}^{N-1}\frac{1-Z_q}{2}
\label{eq:N}
\end{equation}
where $Z_q$ is the Pauli Z gate corresponding to qubit $q$.  Therefore, each substate is composed of a definite number of particles (i.e., qubits in the $|1\rangle$ state) such that, for example, an overall $r=6$ AGP wavefunction
\begin{align}
|\Psi_{\rm AGP}^{6,\rm{full}}\rangle=\frac{1}{2\sqrt{2}}[|000000\rangle+|000011\rangle+|001100\rangle+|110000\rangle
\nonumber\\
+|111100\rangle+|110011\rangle+|001111\rangle+|111111\rangle]
\end{align}
which has contributions from basis states with $N=0$, $N=2$, $N=4$, and $N=6$ particles can be decomposed into the following substates that are eigenfunctions of the number operator and hence have a definite number of particles:
\begin{equation}
|\Psi_{\rm AGP}^{6,0}\rangle=|000000\rangle,
\end{equation}
\begin{equation}
|\Psi_{\rm AGP}^{6,2}\rangle=\frac{1}{\sqrt{3}}[|000011\rangle+|001100\rangle+|110000\rangle],
\end{equation}
\begin{equation}
|\Psi_{\rm AGP}^{6,4}\rangle=\frac{1}{\sqrt{3}}[|111100\rangle+|110011\rangle+|001111\rangle],
\end{equation}
and
\begin{equation}
|\Psi_{\rm AGP}^{6,6}\rangle=|111111\rangle.
\end{equation}
\noindent In general, the overall AGP wavefunction is constructed from its number-conserving substates according to
\begin{equation}
|\Psi_{\rm AGP}^{r,\rm{full}}\rangle=\left(\frac{1}{\sqrt{2}}\right)^{r/2}\left[\sum\limits_{i=0}^{r/2}\sqrt{{r/2 \choose i}}|\Psi_{\rm AGP}^{r,2i}\rangle\right]
\label{eq:subtofull}
\end{equation}

The density matrices associated with the number-conserving substates (${}^{2}D_{r,N}$) are able to be isolated as shown in the following sections such that the signature of ODLRO, $\lambda_D$, can be attained for subblocks of the 2-RDM specific to each substate in addition to the signature of the overall non-particle-conserving AGP state with contributions from each of the number-conserving substates.

\subsubsection{Tomography via Pauli Expectation Values}
\label{sec:tom_of_sub_from_P}
As can be extrapolated from Eq. (\ref{eq:subtofull}), the full density matrix (${}^{2}D_{r,\rm{full}}$) obtained as shown using tomography via Pauli expectation values in \ref{sec:tomviaP1} can be represented as a sum of the density matrices for each of the individual substates (${}^{2}D_{r,N}$) with elements given according to
\begin{gather}
{{}^{2}D_{r,\rm{full}}}^{i,j}_{k,l}=\left(\frac{1}{2}\right)^{r/2}\left[\sum\limits_{i=0}^{r/2}{r/2 \choose i}{{}^{2}D_{r,2i}}^{i,j}_{k,l}\right]
\nonumber\\
=\langle \Psi_{r,\rm{full}}|\hat{a}^\dagger_i\hat{a}^\dagger_j\hat{a}_l\hat{a}_k|\Psi_{r,\rm{full}}\rangle
\label{eq:subtofullD2}
\end{gather}

This relationship alone is insufficient to solve for the number-conserving density matrices in terms of the full density matrix.  Instead, we introduce the number operator from Eq. (\ref{eq:N}) to define a new matrix with elements given by
\begin{equation}
{P_{1}}^{i,j}_{k,l}=\langle \Psi|\hat{a}^{\dagger}_i\hat{a}^{\dagger}_j\hat{a}_l\hat{a}_k\hat{N}|\Psi\rangle
\label{eq:P1intro}
\end{equation}
The elements of this new matrix can be obtained by again directly computing the expectation values of Pauli matrices in a manner directly analogous to that described in \ref{sec:tomviaP1} with the creation and annihilation operators in the basis of Pauli matrices being written as shown in Eqs. (\ref{eq:creation}) and (\ref{eq:annihilation}) and the number operator being written as shown in Eq. (\ref{eq:N}).  Thus, each element can be expressed as
\begin{widetext}
\begin{gather}
\hat{a}^\dagger_j\hat{a}^\dagger_{j+1}\hat{a}_{k+1}\hat{a}_k\hat{N}=\frac{1}{32}\sum\limits_{q=0}^{r}\left(X_j-iY_j\right)\left(X_{j+1}-iY_{j+1}\right)\left(X_{k+1}+iY_{k+1}\right)\left(X_k+iY_k\right)\left(1-Z_q\right)
\end{gather}
\end{widetext}
so that the expectation values of four- and five-qubit Pauli expressions---such as $X_jY_{j+1}X_{k+1}Y_k$ and $X_jY_{j+1}X_{k+1}Y_kZ_q$---can be directly probed on the quantum device in order to compute the each element of the $P_1$ matrix.

Each element of the $P_1$ matrix can additionally be represented as the following linear combination of number-conserving density matrices (${}^{2}D_{r,N}$)
\begin{equation}
{P_1}^{i,j}_{k,l}=\left(\frac{1}{2}\right)^{r/2}\left[\sum\limits_{i=0}^{r/2}{r/2 \choose i}\left[(2i)^1\right]\left({{}^{2}D_{r,2i}}^{i,j}_{k,l}\right)\right]
\label{eq:P1sub}
\end{equation}

\noindent Similarly, other matrices can be defined with elements
\begin{equation}
{P_{x}}^{i,j}_{k,l}=\langle \Psi|\hat{a}^{\dagger}_i\hat{a}^{\dagger}_j\hat{a}_l\hat{a}_k\hat{N}^x|\Psi\rangle
\label{eq:P_xintro}
\end{equation}
which can be additionally be represented according to
\begin{equation}
{P_x}^{i,j}_{k,l}=\left(\frac{1}{2}\right)^{r/2}\left[\sum\limits_{i=0}^{r/2}{r/2 \choose i}\left[(2i)^x\right]\left({{}^{2}D_{r,2i}}^{i,j}_{k,l}\right)\right]
\label{eq:Pxsub}
\end{equation}

Obtaining ${}^2D_{r,\rm{full}}$, $P_1$, $P_2$, etc. directly from probing expectation values of Pauli matrices and solving the system of linear equations described by Eqs. (\ref{eq:subtofullD2}), (\ref{eq:P1sub}), and (\ref{eq:Pxsub}) for the individual number-conserving density matrices (${}^{2}D_{r,N}$) allows for the computation of the signature of ODLRO ($\lambda_D$) corresponding to each of the number-conserving substates.

As can be readily observed, this methodology for obtaining the number-conserving substates quickly becomes prohibitively expensive as system size is increased.  As such, this study relies on the methodology presented in the following section.


\subsubsection{Tomography via Direct Computation of the Wavefunction}
Wavefunctions corresponding to each individual number-conserving substate can be determined from the probability information obtained from a quantum device in a manner analogous to how the full AGP wavefunction is prepared as described in \ref{sec:tompsi}.  Specifically, for a specific possible, even value of $N \in \{0, 2, 4,\dots,r\}$, each element of the number-conserving wavefunction ($|\Psi_i^{r,N}\rangle$) in the basis of the $2^r$ possible stated ($|\eta_i\rangle$) can be obtained according to
\begin{equation}
|\Psi_i^{r,N}\rangle=\begin{cases} 0 & if \ \hat{N}|\eta_i\rangle\ne N |\eta_i\rangle\\
+|\langle \eta_i | \Psi \rangle| & if \ \hat{N}|\eta_i\rangle=N |\eta_i\rangle
\end{cases}
\end{equation}
whereby if the basis state $|\eta_i\rangle$ contains $N$ particles (i.e., has $N$ qubits in the $|1\rangle$ state), the element of the number-conserving wavefunction ($|\Psi_i^{r,N}\rangle$) corresponding to that basis state is identical to that from the full wavefunction ($|\Psi_i\rangle$); however, if the basis state $|\eta_i\rangle$ doesn't contain $N$ particles, $|\Psi_i^{r,N}\rangle$ is set to zero.  The resulting number-conserving wavefunction is then normalized to one.

The signature of ODLRO ($\lambda_D$) can then be obtained directly from the number-conserving wavefunction in the manner described in \ref{sec:tompsi}.  Note that this projection can act as a form of error mitigation as contributions from bases corresponding to odd-numbered basis states---and indeed all basis states not corresponding to the number of particles of interest---are omitted.  Additionally note that for QASM simulation, this methodology for isolating the number-conserving substates from the data obtained from the quantum device yields identical results within sampling error to the methodology presented in \ref{sec:tom_of_sub_from_P} while being significantly less computationally expensive. As such, tomography via direct computation of the wavefunction is employed in this study.

\color{black}
\subsection{Description of Noise on Near-Term Quantum Devices}
\label{sec:noise}
Three main classes of errors lead to the deviation of physical qubits from the idealized logical qubits: namely, gate noise, readout noise, and decoherence. Quantum gate noise/error refers to a situation where the application of a unitary gate $\hat{U}$ to a quantum state $|\Psi\rangle$ yields a result that deviates from $\hat{U}|\Psi\rangle$. This class of error is caused by either imprecisely calibrated control of the qubits and/or imperfect isolation of qubits from their environment, and the overall gate error increases with the number of gates applied. Readout noise/error refers to transmission line noise that makes the $|0\rangle$ state appear to be $|1\rangle$ or vice versa; it can be caused by the probability distributions of the measured physical quantities that correspond to the $|0\rangle$ and $|1\rangle$ states overlapping and/or the qubit decaying during readout. Decoherence involves interactions with external systems (vibrations, temperature fluctuations, electromagnetic waves, etc.) leading to the degradation of the quantum state prepared on quantum devices. As both the number of gates applied to a system---and hence gate noise---and decoherence tend to increase with system size ($r$), larger-scale quantum computations often involve more and more error \cite{preskill_2018}. See the Supplemental Information of Ref.~\onlinecite{Sager_2020} for a more thorough exploration of error on near-term quantum devices.

\subsection{Analysis of Errors Via Joint Probabilities of Occupation}
Comparing the \color{black} results \color{black} obtained from the simulated and experimental Melbourne data illustrates the error associated with the Melbourne device. The probability of a given  orbital (i.e. qubit) being occupied if the orbital with index 0 (i.e., Q0) is occupied was computed for all combinations of total orbitals (qubits, $r$) and particles ($|1\rangle$ qubit states, $N$), and the results for $N=2$ are shown in Tab. \ref{tab:jointprob} in order to comment on relative error based on system size ($r$).
\color{black} Note that all other joint probabilities for Melbourne and  QASM  simulation are given in Tabs. \ref{tab:mel} and \ref{tab:sim}, respectively.  \color{black}

As can be seen from the joint probability data in Tab. \ref{tab:jointprob}, simulated results exactly demonstrate the orbital (qubit) pairing that we program into the system as the only possible \color{black} two-particle \color{black} orientation with Q0 being occupied should be to have Q1 simultaneously occupied.  However, due to errors on Melbourne, this ideal behavior is not exactly recreated on the experimental quantum device.  Specifically, the joint probability of occupying Q0 and Q1 is not unity and seems to decrease with increasing system size.  Additionally, the joint probability of Q0 and Qi where $i\ne 0,1$ should be zero as is seen in \color{black} QASM \color{black} simulation; however, the experimental data demonstrates that other double-excitations are contributing to the overall two-particle substate, indicating error in either state preparation or measurement.  Overall, the error associated with noisy near-term quantum computers decreases the signature of \color{black} ODLRO\color{black}, indicating less \color{black} ODLRO \color{black} character for the experimentally-prepared states than predicted by \color{black} QASM \color{black} simulation.  In order to best construct and probe \color{black} entangled \color{black} states on quantum computers, then, errors on real-world devices need to be minimized.

\begin{table*}[tbh!]
\begin{tabular}{|c|c|c|}
\hline
r,N& \(\displaystyle \lambda_D^{sim.} \) & Probability of Occupation if Particle on Q0 (sim.)  \\ \hline
2, 2               & 0.502                               & {[}x, 1.000{]}                                                         \\
4, 2               & 0.745                               & {[}x, 1.000, 0.000, 0.000{]}                                    \\
6, 2               & 0.996                               & {[}x, 1.000, 0.000, 0.000, 0.000, 0.000{]}               \\
8, 2               & 1.253                               & {[}x, 1.000, 0.000, 0.000, 0.000, 0.000, 0.000, 0.000{]}   \\
10, 2              & 1.500                               & {[}x, 1.000, 0.000, 0.000, 0.000, 0.000, 0.000, 0.000, 0.000, 0.000{]}   \\
12, 2              & 1.755                               & {[}x, 1.000, 0.000, 0.000, 0.000, 0.000, 0.000, 0.000, 0.000, 0.000, 0.000, 0.000{]}  \\
14, 2              & 1.991                               & {[}x, 1.000, 0.000, 0.000, 0.000, 0.000, 0.000, 0.000, 0.000, 0.000, 0.000, 0.000, 0.000, 0.000{]} \\ \hline

r,N& \(\displaystyle \lambda_D^{mel.} \) & Probability of Occupation if Particle on Q0 (mel.) \\ \hline
2, 2 & 0.437                              & {[}x, 1.000{]} \\
4,2  & 0.616                              & {[}x, 0.972, 0.014, 0.014{]}  \\
6,2  & 0.783                              & {[}x, 0.946, 0.013, 0.010, 0.013, 0.018{]} \\
8,2  & 0.898                              & {[}x, 0.831, 0.016, 0.039, 0.027, 0.031, 0.037, 0.018{]} \\
10,2&0.937                              & {[}x, 0.856, 0.024, 0.011, 0.011, 0.019, 0.017, 0.024, 0.010, 0.027{]}\\
12,2&1.116                              & {[}x, 0.678, 0.067, 0.019, 0.036, 0.026, 0.046, 0.041, 0.026, 0.031, 0.012, 0.017{]}\\
14,2&1.261                              & {[}x, 0.784, 0.026, 0.009, 0.009, 0.009, 0.004, 0.043, 0.013, 0.022, 0.039, 0.030, 0.009, 0.004{]} \\\hline
\end{tabular}
\caption{\label{tab:jointprob}The joint probability of the occupation numbers of other orbitals (qubits) if the first orbital ($Q0$) is filled for \color{black} QASM \color{black} simulation (sim.) and ibmq\_16\_melbourne results.}
\end{table*}

\subsection{Experimental \color{black} QASM Simulator and \color{black} Quantum Device Specifications}
\label{sec:devicedetails}
Throughout this work, we have employed the ibmq\_qasm\_simulator \cite{QASM} and the ibmq\_16\_melbourne \cite{melbourne} IBM Quantum Experience device, which \color{black} are available online.  The QASM simulator is  a general-purpose simulator that emulates execution of quantum circuits in either an ideal manner (i.e., with only sampling error) or subject to highly-configurable noise modeling; in this study, all reported QASM results are ideal.    The ibmq\_16\_melbourne \color{black} device is composed of fixed-frequency transmon qubits with co-planer waveguide resonators \cite{koch_2007,chow_2011}. Experimental calibration data and connectivity for this device---as well as that for other devices employed in obtaining supplemental data---is \color{black} given in Tablels \ref{yorktown}-\ref{rochester}.

\subsection{State Preparation Fidelity}
To provide a metric on which to judge the degree to which the expected state preparation was prepared on the quantum devices employed, we include the state preparation fidelity given by \cite{chuang_2010}
\begin{equation}
 \mathcal{F}(\Psi_{\rm ideal},\Psi_{\rm exp.})= |\langle \Psi_{\rm ideal}|\Psi_{\rm exp.}\rangle|^2
\end{equation}
where $|\Psi_{\rm ideal}\rangle$ is the wavefunction corresponding to the result of applying the unitary obtained from the matrix form of the state preparation given in Eq. (\ref{eq:overallprep}) to the all-zero initial state $\left(\bigotimes\limits_{p=0}^{r-1}|0\rangle\right)$ and is hence the ideal expected outcome for a given state preparation on a device with no error and where $|\Psi_{\rm exp.}\rangle$ represents the wavefunction obtained from the quantum device.  The state preparation fidelities for ibmq\_16\_melbourne and QASM simulation are reported in Tabs. \ref{tab:mel} and \ref{tab:sim}.

\subsection{Additional Device Data}
While only data from QASM simulation and the ibmq\_16\_melbourne quantum device are presented, additional data for ibmq\_5\_yorktown, ibmq\_santiago, and ibmq\_rochester are provided in Tabs. \ref{tab:yorktown}-\ref{tab:rochester}.
\color{black}

\begin{table*}[tbh!]
\begin{tabular}{c|c|c|c|c}
\caption{\label{tab:mel} All eigenvalue ($\lambda_D$) information for the non-number-conserving overall state (all) and the number-conserving substates are given with state preparation fidelities ($\mathcal{F}$) and joint probability of occupation numbers of other orbitals (qubits) if the first orbital ($Q0$) is filled for ibmq\_16\_melbourne.}
r  & N   & \(\displaystyle \lambda_D \) & \(\displaystyle \mathcal{F} \) & Probability of Occupation if Particle in Q0 \\\hline
   & all & 0.437     & 0.924        & {[}x, 0.930{]}                                                                                     \\
2  & 0   & 0.000     & 1.000        & N/a                                                                                                        \\
   & 2   & 1.000     & 1.000        & {[}x, 1.000{]}                                                                                     \\\hline
   & all & 0.616     & 0.788        & {[}x, 0.92, 0.476, 0.472{]}                                                                        \\
4   & 0   & 0.000     & 1.000        & N/a                                                                                                        \\
  & 2   & 0.971     & 0.971        & {[}x, 0.972, 0.014, 0.014{]}                                                                       \\
   & 4   & 1.000     & 1.000        & {[}x, 1.000, 1.000, 1.000{]}                                                                       \\\hline
   & all & 0.783     & 0.658        & {[}x, 0.920, 0.484, 0.483, 0.424, 0.438{]}                                                         \\
   & 0   & 0.000     & 1.000        & N/a                                                                                                        \\
6  & 2   & 0.923     & 0.916        & {[}x, 0.946, 0.013, 0.010, 0.013, 0.018{]}                                                         \\
   & 4   & 1.250     & 0.916        & {[}x, 0.974, 0.569, 0.563, 0.444, 0.449{]}                                                         \\
   & 6   & 1.000     & 1.000        & {[}x, 1.000, 1.000, 1.000, 1.000, 1.000{]}                                                         \\\hline
   & all & 0.898     & 0.466        & {[}x, 0.842, 0.436, 0.476, 0.498, 0.487, 0.549, 0.494{]}                                           \\
   & 0   & 0.000     & 1.000        & N/a                                                                                                        \\
8  & 2   & 0.788     & 0.777        & {[}x, 0.831, 0.016, 0.039, 0.027, 0.031, 0.037, 0.018{]}                                           \\
   & 4   & 1.230     & 0.728        & {[}x, 0.871, 0.311, 0.348, 0.351, 0.341, 0.410, 0.368{]}                                           \\
   & 6   & 1.291     & 0.791        & {[}x, 0.928, 0.599, 0.630, 0.698, 0.692, 0.743, 0.710{]}                                           \\
   & 8   & 1.000     & 1.000        & {[}x, 1.000, 1.000, 1.000, 1.000, 1.000, 1.000, 1.000{]}                                           \\\hline
   & all & 0.937     & 0.303        & {[}x, 0.897, 0.475, 0.401, 0.452, 0.541, 0.483, 0.448, 0.384, 0.462{]}                             \\
   & 0   & 0.000     & 1.000        & N/a                                                                                                        \\
   & 2   & 0.631     & 0.624        & {[}x, 0.856, 0.024, 0.011, 0.011, 0.019, 0.017, 0.024, 0.010, 0.027{]}                             \\
10 & 4   & 1.084     & 0.500        & {[}x, 0.898, 0.282, 0.217, 0.241, 0.332, 0.294, 0.263, 0.193, 0.281{]}                             \\
   & 6   & 1.280     & 0.498        & {[}x, 0.928, 0.525, 0.446, 0.517, 0.624, 0.541, 0.495, 0.422, 0.502{]}                             \\
   & 8   & 1.211     & 0.606        & {[}x, 0.958, 0.761, 0.697, 0.776, 0.860, 0.784, 0.740, 0.682, 0.742{]}                             \\
   & 10  & 1.000     & 1.000        & {[}x, 1.000, 1.000, 1.000, 1.000, 1.000, 1.000, 1.000, 1.000, 1.000{]}                             \\\hline
   & all & 1.116     & 0.269        & {[}x, 0.820, 0.486, 0.417, 0.479, 0.512, 0.558, 0.482, 0.420, 0.491, 0.502, 0.490{]}               \\
   & 0   & 0.000     & 1.000        & N/a                                                                                                        \\
   & 2   & 0.647     & 0.626        & {[}x, 0.678, 0.067, 0.019, 0.036, 0.026, 0.046, 0.041, 0.026, 0.031, 0.012, 0.017{]}               \\
12 & 4   & 1.130     & 0.478        & {[}x, 0.763, 0.237, 0.156, 0.213, 0.244, 0.295, 0.236, 0.171, 0.239, 0.229, 0.216{]}               \\
   & 6   & 1.404     & 0.445        & {[}x, 0.840, 0.407, 0.327, 0.425, 0.453, 0.507, 0.432, 0.347, 0.418, 0.428, 0.416{]}                 \\
   & 8   & 1.450     & 0.485        & {[}x, 0.889, 0.600, 0.533, 0.619, 0.650, 0.688, 0.597, 0.531, 0.616, 0.644, 0.633{]}                 \\
   & 10  & 1.288     & 0.616        & {[}x, 0.944, 0.805, 0.763, 0.790, 0.821, 0.868, 0.796, 0.767, 0.820, 0.811, 0.817{]}                 \\
   & 12  & 1.000     & 1.000        & {[}x, 1.000, 1.000, 1.000, 1.000, 1.000, 1.000, 1.000, 1.000, 1.000, 1.000, 1.000{]}               \\\hline
   & all & 1.261     & 0.193        & {[}x, 0.893, 0.389, 0.320, 0.588, 0.491, 0.507, 0.599, 0.499, 0.514, 0.548, 0.538, 0.459, 0.464{]} \\
   & 0   & 0.000     & 1.000        & N/a                                                                                                        \\
   & 2   & 0.555     & 0.535        & {[}x, 0.784, 0.026, 0.009, 0.009, 0.009, 0.004, 0.043, 0.013, 0.022, 0.039, 0.030, 0.009, 0.004{]} \\
   & 4   & 1.038     & 0.375        & {[}x, 0.840, 0.100, 0.069, 0.205, 0.135, 0.175, 0.273, 0.211, 0.215, 0.241, 0.252, 0.139, 0.146{]} \\
14 & 6   & 1.358     & 0.326        & {[}x, 0.869, 0.229, 0.167, 0.440, 0.310, 0.351, 0.478, 0.360, 0.380, 0.418, 0.413, 0.287, 0.300{]} \\
   & 8   & 1.510     & 0.335        & {[}x, 0.906, 0.382, 0.302, 0.616, 0.506, 0.529, 0.641, 0.510, 0.519, 0.575, 0.563, 0.472, 0.478{]}   \\
   & 10  & 1.483     & 0.402        & {[}x, 0.933, 0.576, 0.497, 0.794, 0.715, 0.699, 0.760, 0.665, 0.683, 0.700, 0.688, 0.648, 0.640{]}   \\
   & 12  & 1.282     & 0.561        & {[}x, 0.963, 0.789, 0.741, 0.907, 0.879, 0.864, 0.875, 0.805, 0.820, 0.829, 0.829, 0.843, 0.856{]} \\
   & 14  & 1.000     & 1.000        & {[}x, 1.000, 1.000, 1.000, 1.000, 1.000, 1.000, 1.000, 1.000, 1.000, 1.000, 1.000, 1.000, 1.000{]}
\end{tabular}
\end{table*}

\begin{table*}[tbh!]
\begin{tabular}{c|c|c|c|c}
\caption{\label{tab:sim} All eigenvalue ($\lambda_D$) information for the non-number-conserving overall state (all) and the number-conserving substates are given with state preparation fidelities ($\mathcal{F}$) and joint probability of occupation numbers of other orbitals (qubits) if the first orbital ($Q0$) is filled for QASM simulation.}
r  & N   & \(\displaystyle \lambda_D \) & \(\displaystyle \mathcal{F} \) & Probability of Occupation if Particle in Q0 \\\hline
   & all & 0.502     & 1.000        & {[}x, 1.000{]}                                                                                     \\
2  & 0   & 0.000     & 1.000        & N/a                                                                                                        \\
   & 2   & 1.000     & 1.000        & {[}x, 1.000{]}                                                                                     \\\hline
   & all & 0.745     & 1.000        & {[}x, 1.000, 0.506, 0.506{]}                                                                       \\
4   & 0   & 0.000     & 1.000        & N/a                                                                                                        \\
  & 2   & 1.000     & 1.000        & {[}x, 1.000, 0.000, 0.000{]}                                                                       \\
   & 4   & 1.000     & 1.000        & {[}x, 1.000, 1.000, 1.000{]}                                                                       \\\hline
   & all & 0.996     & 1.000        & {[}x, 1.000, 0.491, 0.491, 0.497, 0.497{]}                                                         \\
   & 0   & 0.000     & 1.000        & N/a                                                                                                        \\
6  & 2   & 1.000     & 1.000        & {[}x, 1.000, 0.000, 0.000, 0.000, 0.000{]}                                                         \\
   & 4   & 1.333     & 1.000        & {[}x, 1.000, 0.494, 0.494, 0.506, 0.506{]}                                                         \\
   & 6   & 1.000     & 1.000        & {[}x, 1.000, 1.000, 1.000, 1.000, 1.000{]}                                                         \\\hline
   & all & 1.253     & 1.000        & {[}x, 1.000, 0.505, 0.505, 0.498, 0.498, 0.506, 0.506{]}                                           \\
   & 0   & 0.000     & 1.000        & N/a                                                                                                        \\
8  & 2   & 1.000     & 1.000        & {[}x, 1.000, 0.000, 0.000, 0.000, 0.000, 0.000, 0.000{]}                                           \\
   & 4   & 1.499     & 1.000        & {[}x, 1.000, 0.334, 0.334, 0.331, 0.331, 0.335, 0.3358{]}                                          \\
   & 6   & 1.500     & 1.000        & {[}x, 1.000, 0.671, 0.671, 0.657, 0.657, 0.673, 0.673{]}                                           \\
   & 8   & 1.000     & 1.000        & {[}x, 1.000, 1.000, 1.000, 1.000, 1.000, 1.000, 1.000{]}                                           \\\hline
   & all & 1.500     & 1.000        & {[}x, 1.000, 0.499, 0.499, 0.499, 0.499, 0.502, 0.502, 0.490, 0.490{]}                             \\
   & 0   & 0.000     & 1.000        & N/a                                                                                                        \\
   & 2   & 1.000     & 1.000        & {[}x, 1.000, 0.000, 0.000, 0.000, 0.000, 0.000, 0.000, 0.000, 0.000{]}                             \\
10 & 4   & 1.598     & 1.000        & {[}x, 1.000, 0.233, 0.233, 0.274, 0.274, 0.263, 0.263, 0.230, 0.230{]}                             \\
   & 6   & 1.798     & 1.000        & {[}x, 1.000, 0.519, 0.519, 0.481, 0.481, 0.499, 0.499, 0.500, 0.500{]}                             \\
   & 8   & 1.600     & 1.000        & {[}x, 1.000, 0.743, 0.743, 0.760, 0.760, 0.758, 0.758, 0.739, 0.739{]}                             \\
   & 10  & 1.000     & 1.000        & {[}x, 1.000, 1.000, 1.000, 1.000, 1.000, 1.000, 1.000, 1.000, 1.000{]}                             \\\hline
   & all & 1.755     & 1.000        & {[}x, 1.000, 0.518, 0.518, 0.499, 0.499, 0.508, 0.508, 0.502, 0.502, 0.509, 0.509{]}               \\
   & 0   & 0.000     & 1.000        & N/a                                                                                                        \\
   & 2   & 1.000     & 1.000        & {[}x, 1.000, 0.000, 0.000, 0.000, 0.000, 0.000, 0.000, 0.000, 0.000, 0.000, 0.000{]}               \\
12 & 4   & 1.664     & 1.000        & {[}x, 1.000, 0.228, 0.228, 0.205, 0.205, 0.184, 0.184, 0.181, 0.181, 0.202, 0.202{]}               \\
   & 6   & 1.997     & 1.000        & {[}x, 1.000, 0.410, 0.410, 0.376, 0.376, 0.389, 0.389, 0.432, 0.432, 0.393, 0.393{]}               \\
   & 8   & 1.997     & 1.000           & {[}x, 1.000, 0.615, 0.615, 0.590, 0.590, 0.619, 0.619, 0.567, 0.567, 0.609, 0.609{]}               \\
   & 10  & 1.666     & 1.000        & {[}x, 1.000, 0.792, 0.792, 0.810, 0.810, 0.802, 0.802, 0.793, 0.793, 0.802, 0.802{]}               \\
   & 12  & 1.000     & 1.000        & {[}x, 1.000, 1.000, 1.000, 1.000, 1.000, 1.000, 1.000, 1.000, 1.000, 1.000, 1.000{]}               \\\hline
   & all & 1.991     & 1.000        & {[}x, 1.000, 0.494, 0.494, 0.497, 0.497, 0.501, 0.501, 0.506, 0.506, 0.500, 0.500, 0.494, 0.494{]} \\
   & 0   & 0.000     & 1.000        & N/a                                                                                                        \\
   & 2   & 1.000     & 0.999        & {[}x, 1.000, 0.000, 0.000, 0.000, 0.000, 0.000, 0.000, 0.000, 0.000, 0.000, 0.000, 0.000, 0.000{]} \\
   & 4   & 1.707     & 1.000        & {[}x, 1.000, 0.134, 0.134, 0.167, 0.167, 0.149, 0.149, 0.205, 0.205, 0.167, 0.167, 0.177, 0.177{]} \\
14 & 6   & 2.137     & 1.000        & {[}x, 1.000, 0.334, 0.334, 0.332, 0.332, 0.315, 0.315, 0.330, 0.330, 0.359, 0.359, 0.331, 0.331{]} \\
   & 8   & 2.278     & 1.000        & {[}x, 1.000, 0.508, 0.508, 0.477, 0.477, 0.514, 0.514, 0.518, 0.518, 0.482, 0.482, 0.502, 0.502{]} \\
   & 10  & 2.138     & 1.000        & {[}x, 1.000, 0.649, 0.649, 0.679, 0.679, 0.684, 0.684, 0.669, 0.669, 0.675, 0.675, 0.644, 0.644{]} \\
   & 12  & 1.713     & 1.000        & {[}x, 1.000, 0.835, 0.835, 0.864, 0.864, 0.835, 0.835, 0.818, 0.818, 0.813, 0.813, 0.835, 0.835{]} \\
   & 14  & 1.000     & 1.000        & {[}x, 1.000, 1.000, 1.000, 1.000, 1.000, 1.000, 1.000, 1.000, 1.000, 1.000, 1.000, 1.000, 1.000{]}
\end{tabular}
\end{table*}

\begin{table*}[tbh!]
\centering
\caption{\label{yorktown}Calibration data for ``Yorktown''}
\begin{tabular}{cccccccc}
\hline\hline

\textbf{Device:} & ibmqx2 (``Yorktown'') \\ \hline
\textbf{Calibration Date:} & 09/10/20 \\ \hline
\textbf{Version:} & 2.0.0 \\ \hline
\textbf{Gate time (ns):} & 71.1 \\ \hline\hline
\textbf{Qubit:} && 0 & 1 & 2 & 3 & 4 \\ \hline
& \textit{T2 ($\mu$s)} & 24.6 & 24.8 & 90.8 & 41.5 & 45.0 \\ \hline
& \textit{f (GHz)} & 5.28 & 5.25 & 5.03 & 5.29 & 5.08 \\ \hline
& \textit{T1 ($\mu$s)} & 46.7 & 40.1 & 60.2 & 60.9 & 73.3 \\ \hline
& \textit{Gate Error ($10^{-3}$)} & 0.99 & 1.94 & 0.59 & 0.49 & 0.52 \\ \hline
& \textit{Readout Error ($10^{-3}$)} &  65.2 & 41.6 & 29.2 & 17.0 & 30.8 \\ \hline\hline
\textbf{Multi-Qubit:} && 0,1 & 0,2 & 1,2 & 2,3 & 2,4 & 3,4 \\ \hline
& \textit{Error ($10^{-3}$)} & 20.9 & 14.7 & 22.2 & 15.3 & 15.0 & 13.7\\ \hline
\label{table:cal1}
\end{tabular}
\end{table*}

\begin{table*}[tbh!]
\centering
\caption{\label{santiago}Calibration data for ``Santiago''}
\begin{tabular}{ccccccc}
\hline\hline

\textbf{Device:} & ibmq\_santiago (``Santiago'') \\ \hline
\textbf{Calibration Date:} & 09/10/20 \\ \hline
\textbf{Version:} & 2.0.0 \\ \hline
\textbf{Gate time (ns):} & 561.78 \\ \hline\hline
\textbf{Qubit:} && 0 & 1 & 2 & 3 & 4 \\ \hline
& \textit{T2 ($\mu$s)} &149.2 & 86.3 & 100.8 & 108.1 & 123.2 \\ \hline
& \textit{f (GHz)} & 4.83 & 4.62 & 4.82 & 4.74 & 4.82 \\ \hline
& \textit{T1 ($\mu$s)} & 74.0 & 190.2 & 138.2 & 161.5 & 106.9 \\ \hline
& \textit{Gate Error ($10^{-3}$)} & 0.32 & 0.18 & 0.19 & 0.19 & 0.20 \\ \hline
& \textit{Readout Error ($10^{-3}$)} & 22.6 & 12.1 & 8.8 & 10.9 & 7.5 \\ \hline\hline
\textbf{Multi-Qubit:} && 0,1 & 1,2 & 2,3 & 3,4  &  \\ \hline
& \textit{Error ($10^{-3}$)} & 10.5 & 7.8 & 5.8 & 5.4 &  \\ \hline
\label{table:calsan}
\end{tabular}
\end{table*}

\begin{turnpage}
\begin{table*}[tbh!]
\small
\centering
\caption{\label{melbourne}Calibration data for ``Melbourne''}
\begin{tabular}{cccccccccccccccccccccc}
\hline\hline
\textbf{Device:}           & ibmq\_16\_melbourne (``Melbourne'')                &      &      &       &        &      &      &       &      &       &      &      &       &       &       &      &      &       &       &       &       \\ \hline
\textbf{Calibration Date:} & 09/02/20                                           &      &      &       &        &      &      &       &      &       &      &      &       &       &       &      &      &       &       &       &       \\ \hline
\textbf{Version:}          & 2.0.0                                             &      &      &       &        &      &      &       &      &       &      &      &       &       &       &      &      &       &       &       &       \\ \hline
\textbf{Gate time (ns):}   & 53.3                                              &      &      &       &        &      &      &       &      &       &      &      &       &       &       &      &      &       &       &       &       \\ \hline\hline
\textbf{Qubit:}            &                                                   & 0    & 1    & 2     & 3      & 4    & 5    & 6     & 7    & 8     & 9    & 10   & 11    & 12    & 13    & 14   &      &       &       &       &       \\ \hline
                                            & \textit{T2 ($\mu$s)} & 95.6 & 56.0 & 87.9 & 16.5 & 63.1 & 37.7 & 77.4 & 14.4 & 101.6 & 30.9 & 92.3 & 24.9 & 68.8 & 34.4 & 64.6 &  &  &  &  &        \\ \hline
                                            & \textit{f (GHz)}                 & 5.11 & 5.24 & 5.04 & 4.89 & 5.02 & 5.07 & 4.93 & 4.98 & 4.75 & 4.97 & 4.94 & 5.00 & 4.76 & 4.97 & 5.01 &  &  &  &  &        \\ \hline
                                            & \textit{T1 ($\mu$s)}         & 73.7 & 56.0 & 63.1 & 66.7 & 59.1 & 20.3 & 72.5 & 38.0 & 108.6 & 44.5 & 64.5 & 14.6 & 71.9 & 32.2 & 45.8 &  &  &  &  &       \\ \hline
                                            & \textit{Gate Error ($10^{-3}$)}    & 0.64 & 1.03 & 0.64 & 0.50 & 0.80 & 3.16 & 1.19 & 1.66 & 0.76 & 1.48 & 2.00 & 50.90 & 0.83 & 2.14 & 0.61 &  &  &  &  &        \\ \hline
                                            & \textit{Readout Error ($10^{-3}$)} & 27.8 & 34.3 & 25.2 & 89.1 & 38.7 & 43.9 & 33.1 & 66.0 & 55.4 & 78.3 & 72.9 & 268.3 & 159.3 & 80.1 & 50.7 &  &  &  &  &        \\ \hline\hline
\textbf{Multi-Qubit:}      &                                                   & 0,1  & 0,14 & 1,2   & 1,13   & 2,3  & 2,12 & 3,4   & 3,11 & 4,5   & 4,10 & 5,6  & 5,9   & 6,8   & 7,8   & 8,9  & 9,10 & 10,11 & 11,12 & 12,13 & 13,14 \\ \hline
                                            & \textit{Error ($10^{-3}$)}         & 19.9 & 28.1 & 12.0 & 45.8 & 25.5 & 38.8 & 17.1 & 1000.0 & 26.2 & 27.8 & 58.3 & 32.8 & 28.4 & 34.3 & 39.4 & 39.7 & 1000.0 & 1000.0 & 22.5 & 35.7 \\\hline
\label{table:cal2}
\end{tabular}
\end{table*}


\begin{table*}[tbh!]
\footnotesize
\centering
\caption{\label{rochester}Calibration data for ``Rochester''}
\begin{tabular}{cccccccccccccccccccccc}
\hline\hline
\textbf{Device:}           & ibmq\_rochester (``Rochester'')                &      &      &       &        &      &      &       &      &       &      &      &       &       &       &      &      &       &       &       &       \\ \hline
\textbf{Calibration Date:} & 2/26/20                                           &      &      &       &        &      &      &       &      &       &      &      &       &       &       &      &      &       &       &       &       \\ \hline
\textbf{Version:}          & 1.2.0                                             &      &      &       &        &      &      &       &      &       &      &      &       &       &       &      &      &       &       &       &       \\ \hline
\textbf{Gate time (ns):}   & 53.3                                              &      &      &       &        &      &      &       &      &       &      &      &       &       &       &      &      &       &       &       &       \\ \hline\hline
\textbf{Qubit:}            &                                                   & 0    & 1    & 2     & 3      & 4    & 5    & 6     & 7    & 8     & 9    & 10   & 11    & 12    & 13    & 14   &    15  &   16    &   17    &   18    &   19    \\ \hline
& \textit{T2 ($\mu$s)} & 77.9 & 48.3 & 61.1 & 78.3 & 59.2 & 74.6 & 63.8 & 32.9 & 53.2 & 55.4 & 73.5 & 56.8 & 67.8 & 66.7 & 64.7 & 26.0 & 46.7 & 11.2 & 43.4 & 47.0      \\ \hline
                                      & \textit{f (GHz)}   & 4.92 & 5.08 & 5.00 & 5.05 & 4.94 & 5.05 & 5.05 & 4.93 & 5.03 & 4.97 & 5.06 & 4.93 & 5.05 & 4.95 & 5.07 & 4.87 & 5.08 & 4.99 & 5.06 & 5.02  \\ \hline
                                      & \textit{T1 ($\mu$s)} & 51.0 & 41.8 & 54.2 & 49.3 & 51.9 & 53.0 & 46.6 & 56.4 & 45.8 & 44.3 & 59.4 & 64.6 & 57.3 & 53.9 & 54.9 & 50.5 & 36.6 & 47.6 & 51.2 & 62.6 \\ \hline
                                      & \textit{Gate Error ($10^{-3}$)}    & 0.88 & 1.24 & 4.48 & 3.74 & 1.16 & 1.61 & 1.43 & 13.95 & 8.85 & 1.80 & 0.88 & 1.47 & 1.96 & 2.06 & 0.74 & 0.95 & 0.61 & 1.18 & 1.09 & 1.19  \\ \hline
                                      & \textit{Readout Error ($10^{-3}$)} & 76.9 & 210.6 & 129.4 & 133.1 & 47.5 & 256.3 & 61.9 & 361.2 & 312.5 & 103.8 & 61.9 & 152.5 & 145.0 & 280.6 & 98.8 & 63.7 & 341.9 & 66.3 & 34.4 & 85.6 \\ \hline
          &                                                   & 20    & 21    & 22     & 23      & 24    & 25    & 26     & 27    & 28     & 29    & 30   & 31    & 32    & 33    & 34   &  35    &  36     &  37     &  38     &  39     \\ \hline
& \textit{T2 ($\mu$s)} & 84.6 & 38.0 & 14.5 & 42.3 & 37.8 & 58.2 & 18.1 & 66.6 & 63.3 & 14.7 & 27.8 & 86.6 & 69.7 & 76.2 & 32.0 & 22.6 & 11.5 & 50.7 & 50.7 & 45.4         \\ \hline
                                      & \textit{f (GHz)}                   & 4.89 & 5.14 & 4.90 & 4.96 & 4.88 & 4.97 & 5.02 & 4.99 & 5.02 & 5.00 & 5.13 & 4.96 & 5.15 & 5.21 & 4.97 & 5.05 & 5.03 & 5.17 & 4.91 & 5.05  \\ \hline
                                      & \textit{T1 ($\mu$s)}               & 65.4 & 48.1 & 54.0 & 79.1 & 62.0 & 61.2 & 48.8 & 54.0 & 55.6 & 63.9 & 31.2 & 66.2 & 47.2 & 58.8 & 55.0 & 61.4 & 50.1 & 51.6 & 79.1 & 55.5  \\ \hline
                                      &\textit{Gate Error ($10^{-3}$)}    & 0.64 & 4.13 & 1.12 & 1.54 & 1.25 & 0.69 & 1.73 & 1.92 & 0.85 & 2.05 & 3.77 & 1.12 & 2.59 & 1.10 & 1.11 & 2.95 & 4.21 & 1.07 & 0.96 & 1.06  \\ \hline
                                      & \textit{Readout Error ($10^{-3}$)} & 17.5 & 87.5 & 130.0 & 335.0 & 35.6 & 175.0 & 56.3 & 121.3 & 196.3 & 26.2 & 353.8 & 171.3 & 58.1 & 65.0 & 38.7 & 209.4 & 25.6 & 171.3 & 195.6 & 125.0 \\ \hline
         &                                                   &40    & 41    & 42     & 43      & 44    & 45    & 46     & 47    & 48     & 49    & 50   & 51    & 52    &    &    &      &       &       &       &       \\ \hline
& \textit{T2 ($\mu$s)} & 79.6 & 35.3 & 83.7 & 46.3 & 48.1 & 57.0 & 60.9 & 63.8 & 89.4 & 50.2 & 31.1 & 47.8 & 57.9   & & & & & &  &      \\ \hline
                                      & \textit{f (GHz)}                   & 5.06 & 5.05 & 4.99 & 5.07 & 4.98 & 5.09 & 5.03 & 5.07 & 4.97 & 5.07 & 4.96 & 5.13 & 5.04  & & & & & &  & \\ \hline
                                      & \textit{T1 ($\mu$s)}               & 53.4 & 51.0 & 45.2 & 51.7 & 64.4 & 44.5 & 51.8 & 57.2 & 68.4 & 46.5 & 60.9 & 59.5 & 54.5 & & & & & &  & \\ \hline
                                      & \textit{Gate Error ($10^{-3}$)}    & 1.76 & 1.87 & 0.62 & 1.05 & 0.90 & 1.00 & 1.46 & 0.86 & 1.38 & 0.88 & 0.90 & 8.12 & 5.01  & & & & & &  & \\ \hline
                                      &\textit{Readout Error ($10^{-3}$)} & 121.9 & 446.3 & 36.9 & 197.5 & 103.8 & 154.4 & 243.8 & 161.9 & 28.1 & 345.6 & 104.4 & 110.6 & 266.3 & & & & & &  & \\ \hline\hline
\textbf{Multi-Qubit:}      &                                                   & 0,1  & 0,5 & 1,2   & 2,3   & 3,4  & 4,6 & 5,9   & 6,13 & 7,8   & 7,16 & 8,9  & 9,10   & 10,11   & 11,12   & 11,17  & 12,13 & 13,14 & 14,15 & 15,18 & 16,19 \\ \hline
                                            & \textit{Error ($10^{-3}$)}         &  63.0 & 54.1 & 32.9 & 60.5 & 48.4 & 39.2 & 31.6 & 1000.0 & 1000.0 & 1000.0 & 1000.0 & 35.7 & 24.2 & 51.0 & 20.5 & 1000.0 & 46.3 & 80.9 & 35.1 & 22.7 \\\hline
      &                                                   & 17,23  & 18,27  & 19,20   & 20,21   & 21,22  & 21,28 & 22,23   & 23,24 & 24,25   &25,26  & 25,29  & 26,27   &  29,36  & 30,31   & 30,39  & 31,32 & 32,33 & 33,34 & 34,35 & 34,40 \\ \hline
                                            & \textit{Error ($10^{-3}$)}         & 40.1 & 39.4 & 20.8 & 32.8 & 69.4 & 38.2 & 59.2 & 36.4 & 38.8 & 45.2 & 73.1 & 33.6 & 75.4 & 1000.0 & 1000.0 & 39.4 & 31.5 & 37.6 & 38.9 & 33.3 \\\hline
      &                                                   & 35,36  & 36,37 & 37,38   & 38,41   & 39,42  & 40,46 & 41,50   & 42,43 & 43,44   & 44,45 & 44,51  & 45,46   &  46,47  & 47,48   &  48,49 & 48,52 & 49,50 &  &  &  \\ \hline

                                            & \textit{Error ($10^{-3}$)}         & 56.2 & 92.7 & 58.8 & 1000.0 & 38.7 & 40.8 & 1000.0 & 28.0 & 28.0 & 1000.0 & 134.6 & 1000.0 & 29.2 & 19.8 & 31.4 & 49.4 & 29.7 &   &  &  \\\hline\hline
\label{table:cal3}
\end{tabular}
\end{table*}

\end{turnpage}

\begin{table*}[tbh!]
\begin{tabular}{c|c|c|c}
\caption{\label{tab:yorktown} All eigenvalue information for the non-number-conserving overall state (all) and the number-conserving substates are given with joint probability of occupation numbers of other orbitals (qubits) if the first orbital ($Q0$) is filled for ibmq\_5\_yorktown.}
r & N & \(\displaystyle \lambda_D \) & Probability of Occupation if Particle on Q0 \\ \hline
  & all             & 0.460                        & {[}x, 0.902{]}                              \\
2 & 0               & 0.000                        & N/a                                         \\
  & 2               & 1.000                        & {[}x, 1.000{]}                              \\ \hline
  & all             & 0.658                        & {[}x, 0.900, 0.486, 0.493{]}                \\
4 & 0               & 0.000                        & N/a                                         \\
  & 2               & 0.957                        & {[}x, 0.983, 0.006, 0.011{]}                \\
  & 4               & 1.000                        & {[}x, 1.000, 1.000, 1.000{]}           \\
\end{tabular}
\end{table*}

\begin{table*}[tbh!]
\begin{tabular}{c|c|c|c}
\caption{\label{tab:santiago} All eigenvalue information for the non-number-conserving overall state (all) and the number-conserving substates are given with joint probability of occupation numbers of other orbitals (qubits) if the first orbital ($Q0$) is filled for ibmq\_santiago.}
r & N & \(\displaystyle \lambda_D \) & Probability of Occupation if Particle on Q0 \\ \hline
  & all             & 0.429                        & {[}x, 0.959{]}                              \\
2 & 0               & 0.000                        & N/a                                         \\
  & 2               & 1.000                        & {[}x, 1.000{]}                              \\ \hline
  & all             & 0.669                        & {[}x, 0.964, 0.506, 0.508{]}                \\
4 & 0               & 0.000                        & N/a                                         \\
  & 2               & 0.998                        & {[}x, 0.997, 0.001, 0.002{]}                \\
  & 4               & 1.000                        & {[}x, 1.000, 1.000, 1.000{]}
\end{tabular}
\end{table*}

\begin{table*}[tbh!]
\begin{tabular}{c|c|c|c}
\caption{\label{tab:rochester} All eigenvalue information for the non-number-conserving overall state (all) and the number-conserving substates are given with joint probability of occupation numbers of other orbitals (qubits) if the first orbital ($Q0$) is filled for ibmq\_rochester.}
r  & N & \(\displaystyle \lambda_D \) & Probability of Occupation if Particle on Q0                                          \\ \hline
   & all             & 0.339                        & {[}x, 0.690{]}                                                                       \\
2  & 0               & 0.000                        & N/a                                                                                  \\
   & 2               & 1.000                        & {[}x, 1.000{]}                                                                       \\ \hline
   & all             & 0.490                        & {[}x, 0.687, 0.507, 0.511{]}                                                         \\
4  & 0               & 0.000                        & N/a                                                                                  \\
   & 2               & 0.779                        & {[}x, 0.785, 0.114, 0.101{]}                                                         \\
   & 4               & 1.000                        & {[}x, 1.000, 1.000, 1.000{]}                                                         \\ \hline
   & all             & 0.673                        & {[}x, 0.771, 0.514, 0.522, 0.501, 0.497{]}                                           \\
   & 0               & 0.000                        & N/a                                                                                  \\
6  & 2               & 0.727                        & {[}x, 0.720, 0.088, 0.099, 0.042, 0.051{]}                                           \\
   & 4               & 1.071                        & {[}x, 0.853, 0.548, 0.552, 0.526, 0.520{]}                                           \\
   & 6               & 1.000                        & {[}x, 1.000, 1.000, 1.000, 1.000, 1.000{]}                                           \\ \hline
   & all             & 0.746                        & {[}x, 0.751, 0.504, 0.484, 0.469, 0.495, 0.519, 0.497{]}                             \\
   & 0               & 0.000                        & N/a                                                                                  \\
8  & 2               & 0.514                        & {[}x, 0.549, 0.075, 0.076, 0.088, 0.087, 0.076, 0.049{]}                             \\
   & 4               & 0.921                        & {[}x, 0.740, 0.389, 0.374, 0.355, 0.373, 0.399, 0.370{]}                             \\
   & 6               & 1.015                        & {[}x, 0.851, 0.689, 0.668, 0.645, 0.686, 0.734, 0.727{]}                             \\
   & 8               & 1.000                        & {[}x, 1.000, 1.000, 1.000, 1.000, 1.000, 1.000, 1.000{]}                             \\ \hline
   & all             & 0.898                        & {[}x, 0.819, 0.483, 0.536, 0.465, 0.432, 0.508, 0.515, 0.477, 0.455{]}               \\
   & 0               & 0.000                        & N/a                                                                                  \\
   & 2               & 0.515                        & {[}x, 0.608, 0.053, 0.072, 0.045, 0.050, 0.042, 0.050, 0.042, 0.038{]}               \\
10 & 4               & 0.966                        & {[}x, 0.759, 0.293, 0.342, 0.254, 0.237, 0.288, 0.300, 0.279, 0.249{]}               \\
   & 6               & 1.163                        & {[}x, 0.847, 0.518, 0.578, 0.494, 0.446, 0.555, 0.564, 0.514, 0.485{]}               \\
   & 8               & 1.108                        & {[}x, 0.925, 0.751, 0.785, 0.751, 0.707, 0.795, 0.801, 0.752, 0.735{]}               \\
   & 10              & 1.000                        & {[}x, 1.000, 1.000, 1.000, 1.000, 1.000, 1.000, 1.000, 1.000, 1.000{]}               \\ \hline
   & all             & 0.974                        & {[}x, 0.846, 0.518, 0.534, 0.568, 0.555, 0.426, 0.422, 0.469, 0.447, 0.476, 0.474{]} \\
   & 0               & 0.000                        & N/a                                                                                  \\
   & 2               & 0.361                        & {[}x, 0.565, 0.056, 0.040, 0.052, 0.048, 0.044, 0.032, 0.028, 0.016, 0.073, 0.044{]} \\
12 & 4               & 0.831                        & {[}x, 0.743, 0.222, 0.234, 0.314, 0.305, 0.192, 0.202, 0.182, 0.172, 0.215, 0.221{]} \\
   & 6               & 1.103                        & {[}x, 0.833, 0.443, 0.473, 0.519, 0.501, 0.350, 0.354, 0.378, 0.357, 0.394, 0.398{]} \\
   & 8               & 1.165                        & {[}x, 0.898, 0.654, 0.673, 0.671, 0.665, 0.534, 0.521, 0.617, 0.585, 0.592, 0.588{]} \\
   & 10              & 1.024                        & {[}x, 0.946, 0.850, 0.854, 0.823, 0.841, 0.752, 0.733, 0.849, 0.816, 0.763, 0.773{]} \\
   & 12              & 1.000                        & {[}x, 1.000, 1.000, 1.000, 1.000, 1.000, 1.000, 1.000, 1.000, 1.000, 1.000, 1.000{]} \\ \hline
   & all & 0.909 & {[}x, 0.529, 0.561, 0.487, 0.551, 0.545, 0.536, 0.515, 0.432, 0.451, 0.538, 0.539, 0.487, 0.485{]} \\
   & 0   & 0.000 & N/a                                                                                                \\
   & 2   & 0.149 & {[}x, 0.093, 0.155, 0.052, 0.031, 0.000, 0.062, 0.103, 0.103, 0.155, 0.062, 0.103, 0.031, 0.052{]} \\
   & 4   & 0.580 & {[}x, 0.302, 0.311, 0.224, 0.161, 0.155, 0.260, 0.204, 0.235, 0.230, 0.280, 0.276, 0.175, 0.188{]} \\
14 & 6   & 0.905 & {[}x, 0.436, 0.436, 0.367, 0.374, 0.366, 0.405, 0.379, 0.333, 0.354, 0.439, 0.438, 0.338, 0.335{]} \\
   & 8   & 1.036 & {[}x, 0.546, 0.589, 0.517, 0.599, 0.600, 0.573, 0.548, 0.446, 0.449, 0.556, 0.568, 0.504, 0.507{]} \\
   & 10  & 1.032 & {[}x, 0.665, 0.713, 0.635, 0.802, 0.797, 0.711, 0.696, 0.582, 0.604, 0.720, 0.698, 0.691, 0.686{]} \\
   & 12  & 0.945 & {[}x, 0.778, 0.850, 0.823, 0.943, 0.925, 0.873, 0.859, 0.764, 0.778, 0.838, 0.831, 0.866, 0.873{]} \\
   & 14  & 1.000 & {[}0, 1.000, 1.000, 1.000, 1.000, 1.000, 1.000, 1.000, 1.000, 1.000, 1.000, 1.000, 1.000, 1.000{]}\end{tabular}
\end{table*}

\newpage
\bibliography{references,references_2}

\end{document}